\documentclass[10pt,prd,twocolumn,showpacs,amsmath,amssymb,aps,floats,floatfix,nofootinbib]{revtex4-1}%
\usepackage{dcolumn}
\usepackage{bm}
\usepackage{amsmath}
\usepackage{amsfonts}
\usepackage{amssymb}
\usepackage{graphicx}
\usepackage{color}
\usepackage{latexsym}
\usepackage{slashed} % slash mark
\usepackage{subfigure}

\def\cm{\mathrm{cm}} % cm
\def\sec{\mathrm{s}} % s
\def\fb{\mathrm{fb}} % fb
\def\ifb{\mathrm{fb}^{-1}} % fb^-1
\def\TeV{\mathrm{TeV}} % TeV
\def\GeV{\mathrm{GeV}} % GeV
\def\pT{p_\mathrm{T}} % p_T
\def\missE{\slashed E} % missing E

\makeatother

\begin{document}

\title{Detecting interactions between dark matter and photons at high energy $e^+e^-$ colliders}
\author{Zhao-Huan Yu$^1$}
\author{Qi-Shu Yan$^{2,3}$}
\author{Peng-Fei Yin$^1$}
\affiliation{$^1$Key Laboratory of Particle Astrophysics,
Institute of High Energy Physics, Chinese Academy of Sciences,
Beijing 100049, China}
\affiliation{$^2$College of Physics Sciences,
University of Chinese Academy of Sciences,
Beijing 100049, China}
\affiliation{$^3$Center for High Energy Physics, Peking University, Beijing 100871, China}

\begin{abstract}
We investigate the sensitivity to the effective operators describing interactions between dark matter particles and photons at future high energy $e^+e^-$ colliders via the $\gamma+ \slashed{E}$ channel. Such operators could be useful to interpret the potential gamma-ray line signature observed by the Fermi-LAT. We find that these operators can be further tested at $e^+ e^-$ colliders by using either unpolarized or polarized beams. We also derive a general unitarity condition for $2 \to n$ processes and apply it to the dark matter production process $e^+e^-\to\chi\chi\gamma$.
\end{abstract}


\maketitle

\section{Introduction}

According to astrophysical and cosmological observations in recent years,
about a quarter of the energy of our Universe is made up of
non-baryonic dark matter (DM), which is further confirmed
by the recent Planck measurement~\cite{Ade:2013zuv}.
Nonetheless, the nature of DM remains an open question. On the market,
the most attractive DM candidate is the weakly interacting massive particle,
whose mass and interaction strength can naturally explain
the DM relic density. Obviously, detecting the signals of DM particles via non gravitational effects is helpful to reveal the mystery of DM.

At the tree level, DM particles should have no direct coupling to photons, otherwise they should be called as ``luminous matter''. Nevertheless, a pair of DM particles can annihilate into two photons via loop-induced processes, i.e. via $\chi\chi\to\gamma\gamma$,
as shown in Refs.~\cite{Bergstrom:1988fp,Rudaz:1989ij,Bergstrom:1997fh}.
The photon produced via such loop-induced processes is monochromatic and carries the energy of the DM particle mass $m_\chi$. If such photons have large flux and can be captured by detectors, they will be identified as a ``line'' and be distinguished from the continuous astrophysical background spectrum clearly. If such a characteristic line signature is detected, it is the ``smoking gun'' discovery for the DM particles.

Recently, several studies of the Fermi-LAT $\gamma$-ray data have shown that there might exist a monochromatic $\gamma$-ray line near the energy $\sim 130\,\GeV$ from the Galactic center region~\cite{Bringmann:2012vr,Weniger:2012tx,Tempel:2012ey}. If such a monochromatic $\gamma$-ray line is true, it could be interpreted as the result of the DM annihilation into photons in the Galactic Center with a cross section of $\left<\sigma_\mathrm{ann}v\right> \sim 10^{-27}\,\cm^3\,\sec^{-1}$. However, the Fermi-LAT collaboration did not confirm such a $\gamma$-ray line in the latest analysis~\cite{Fermi-LAT:2013uma}. Instead, they set the upper limits on the DM annihilation cross section into photons.

In addition to direct searches for the DM particles scattering off nuclei in underground detectors and indirect searches for DM annihilation/decay products from outer space, $\TeV$-scale colliders provide another independent and complementary approach to search for the DM particles produced via high energy collisions.
Although DM particles almost cannot interact with materials of the general-purpose detectors, it has been pointed out that either the mono-jet channel or the mono-photon associating with a large missing energy ($\missE$) can be a distinctive signature in DM searches at both hadron and electron colliders~\cite{Birkedal:2004xn,Konar:2009ae,Beltran:2010ww,Goodman:2010yf,Bai:2010hh,
Fox:2011fx,Rajaraman:2011wf,Dreiner:2012xm,Chae:2012bq,Bhattacherjee:2012ch,Ding:2012sm,Zhou:2013fla,Nelson:2013pqa,Cheung:2010ua}.

\begin{figure}[!htbp]
\centering
\includegraphics[width=0.7\columnwidth]{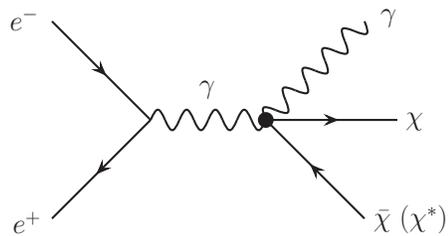}
\caption{Feynman diagram for the mono-photon process
$e^+e^-\to\gamma^* \to \chi\chi\gamma$ at $e^+e^-$ colliders.}
\label{fig:feyn_prod}
\end{figure}

The interaction between DM particles and photons which induces a ``line" signal via the process $\chi\chi\to\gamma\gamma$ can also lead to the process $e^+e^-\to\chi\chi\gamma$ as shown in Fig. \ref{fig:feyn_prod}. Thus the potential $\gamma$-ray line signal can be tested independently at future $e^+e^-$ colliders. In this work, we explore the prospect of the DM searching at $e^+e^-$ colliders by using the mono-photon signature, i.e. the $\gamma+\missE$ channel.

In a model-independent way, the interaction terms between a pair of DM particles and a pair of standard model (SM) particles can be described by effective operators, typically high-dimensional and non-renormalizable~\cite{Beltran:2010ww,Goodman:2010yf,Bai:2010hh,Fox:2011fx,Rajaraman:2011wf,Dreiner:2012xm,Chae:2012bq,Bhattacherjee:2012ch,Beltran:2008xg,
Cao:2009uw,Goodman:2010qn,Cheung:2010ua,Zheng:2010js,Ding:2012sm,Rajaraman:2012fu,Tsai:2012cs,Chen:2013gya,Zhou:2013fla,Nelson:2013pqa}.
Such a treatment would be valid if DM particles couple to SM particles via exchanging some mediators which are sufficiently heavy. In the framework of the effective field theory,
the scattering, annihilation, and production cross sections
of DM particles can be easily related to
each other~\cite{Beltran:2010ww,Goodman:2010yf,Bai:2010hh,Fox:2011fx,Rajaraman:2011wf,Dreiner:2012xm,Chae:2012bq,Bhattacherjee:2012ch,Beltran:2008xg,
Cao:2009uw,Goodman:2010qn,Cheung:2010ua,Zheng:2010js,Ding:2012sm,Rajaraman:2012fu,Tsai:2012cs,Chen:2013gya,Zhou:2013fla,Nelson:2013pqa},
which offers a convenient way to comprehend the correlations among three kinds of DM searching experiments.

In this work, we adopt two effective operators to describe
the interaction between DM particles (either a Dirac or a scalar DM particle) and photons, and investigate their experimental sensitivity at future $e^+e^-$ colliders. We also explore that the effects of beam polarization to the background $e^+e^-\to\nu\bar\nu\gamma$ and the DM signal, and we find that a realistic beam polarization can greatly suppress the background events and enhance the production rate of signal events.

As well-known, the effective theory approach would break down if the collision energy is sufficiently high due to the power dependence on $s$ of matrix elements. In this work, we revisit the method to derive a general unitary condition for $2 \to n$ processes. Thus we apply this novel unitarity condition for the process $e^+e^-\to\chi\chi\gamma$ in order to obtain meaningful bounds.

This paper is organized as follows.
In Sec.~\ref{sec:reach}, we investigate the experimental sensitivity of the $\gamma+\missE$ signature
at future $e^+e^-$ colliders to the effective operators.
We also study the effects of beam polarization in Sec.~\ref{sec:pol}. In Sec.~\ref{sec:uni}, we derive a general unitarity condition for $2 \to n$ processes and use it to check the validity of
our effective operator treatment. We end this work with conclusions and discussions in Sec.~\ref{sec:conc}.

\section{Experimental sensitivity at future $e^+e^-$ colliders \label{sec:reach}}

We consider two types of DM particles: one is a Dirac fermion and the other is a complex scalar particle. We assume a pair of DM particles couple to photons through the following two effective operators:
\begin{equation}
\mathcal{O}_F = \frac{1}{\Lambda^3}
\bar\chi i\gamma_5\chi F_{\mu\nu}{\tilde F}^{\mu\nu}
\label{eq:O_F}
\end{equation}
for the Dirac fermionic DM and
\begin{equation}
\mathcal{O}_S = \frac{1}{\Lambda^2}
\chi^*\chi F_{\mu\nu}F^{\mu\nu}
\label{eq:O_S}
\end{equation}
for the complex scalar DM, respectively.

For nonrelativistic DM particles in the Galaxy, the annihilation cross sections into photons can read
\begin{equation}
\left<\sigma_{\mathrm{ann}}v\right>_{\chi\bar\chi\to 2\gamma}
\simeq \frac{4m_\chi^4}{\pi\Lambda^6},
\end{equation}
and
\begin{equation}
\left<\sigma_{\mathrm{ann}}v\right>_{\chi\chi^*\to 2\gamma}
\simeq \frac{2m_\chi^2}{\pi\Lambda^4}.
\end{equation}
To interpret the Fermi $\gamma$-ray line signal at $\sim 130\,\GeV$, a cross section of $\sim 10^{-27}\,\cm^3\,\sec^{-1}$ is needed, which corresponds to $\Lambda\sim 1\,(3)\,\TeV$ for the fermionic (scalar) DM.

The dimensional analysis tells us that the cross section of the $s$-wave annihilation is proportional to $m_\chi^{2(n-5)} \Lambda^{-2(n-4)}$ for an $n$-dimension operator. The operator for the fermionic DM given in Eq.~\eqref{eq:O_F} has a dimension 7, higher than that for the scalar DM. Therefore the fermionic DM annihilation is more suppressed. As a result, in order to achieve the same cross section for a fixed DM mass, a lower $\Lambda$, which means a stronger coupling, is needed for the fermionic DM.

At the $e^+e^-$ colliders, the leading DM production process for the operators discussed above
is $e^+e^-\to \gamma^* \to \chi\chi\gamma$ which is shown in Fig.~\ref{fig:feyn_prod}. DM particles can pass through the detectors and leave a large missing energy. Therefore, we can use mono-photon and a large missing energy to search the signals.

It is remarkable that the dimensional analysis essentially determine the experimental sensitivity to DM-photon interactions at colliders. In other words, for the operators with the same mass dimension, their collider detection sensitivity are rather similar, while their Lorentz structures play minor roles.
Therefore, we confine to address two representative operators given in Eqs.~\eqref{eq:O_F} and \eqref{eq:O_S}. Interested readers can refer Refs.~\cite{Rajaraman:2012fu,Chen:2013gya} for a more comprehensive discussion about various DM-photon effective operators.

For the $\gamma+\missE$ searching channel,
the SM process $e^+e^-\to\nu\bar\nu\gamma$
is an irreducible background~\cite{Chen:1995yu},
since neutrinos, like DM particles, are also undetectable in a general-purpose detector.
The SM background is contributed by two Feynman diagrams:
one involves the $t$-channel $W$ boson exchange and
the other involves the $s$-channel $Z$ boson exchange.
Another possible background is $e^+e^-\to e^+e^-\gamma$,
where neither the $e^+$ nor the $e^-$ in the final state is detected. Other minor SM backgrounds can be safely  neglected.

We explore the DM searching prospect at the $e^+e^-$ colliders with $\sqrt{s}=250\,\GeV$ (``Higgs factory''), $500\,\GeV$ (typical ILC),
$1\,\TeV$ (upgraded ILC and initial CLIC), and $3\,\TeV$ (ultimate CLIC).
We perform the event simulation with
\texttt{MadGraph\;5}~\cite{Alwall:2011uj},
to which the new particles and couplings are added
through the package \texttt{FeynRules}~\cite{Christensen:2008py}.
The package \texttt{PGS\;4}~\cite{pgs} is used to carry out the fast detector simulation. The energy resolution of the electromagnetic calorimeter is assumed to be
\begin{equation}
\frac{\Delta E}{E} = \frac{16.6\%}{\sqrt{E/\GeV}} \oplus 1.1\%,
\end{equation}
as specified in the design of the ILD detector~\cite{Abe:2010aa}.

We propose the following cuts to suppress the SM background events:
\begin{figure*}[!htbp]
\centering
\subfigure[]{\includegraphics[width=0.9\columnwidth]{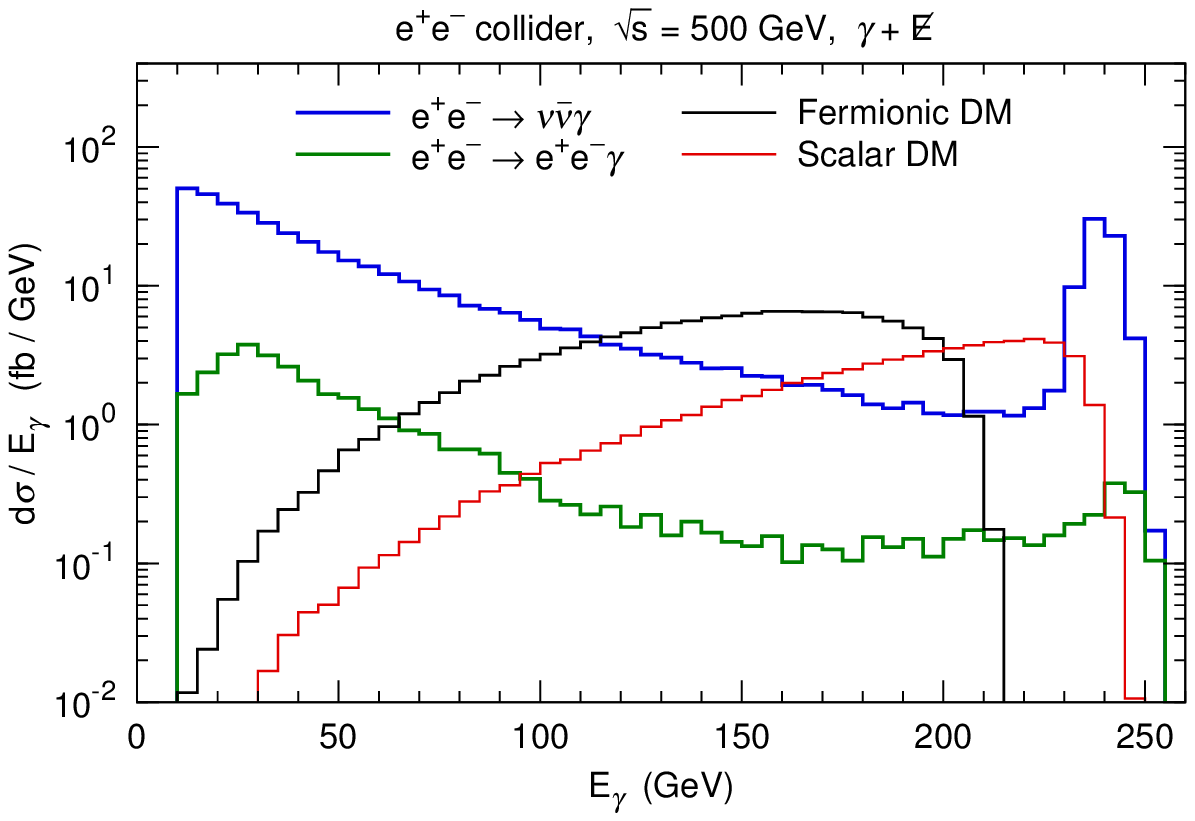}}
\subfigure[]{\includegraphics[width=0.9\columnwidth]{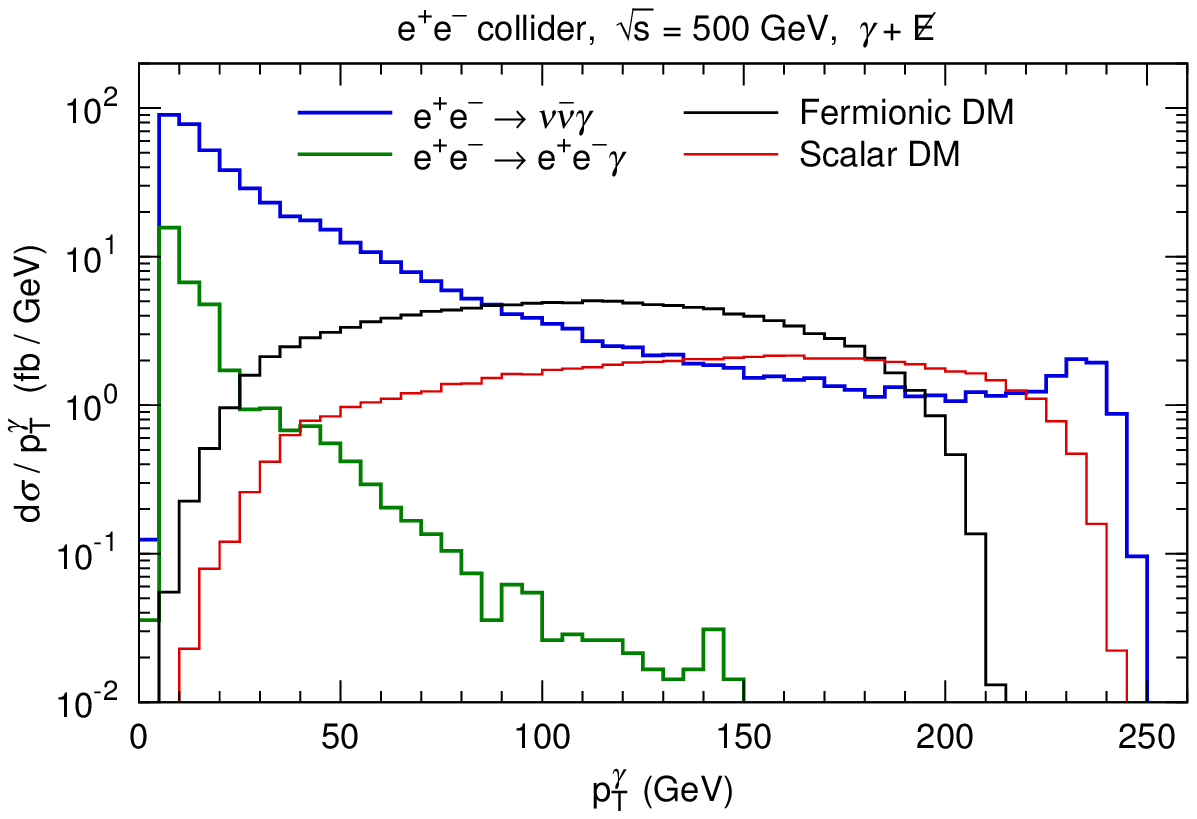}}
\subfigure[]{\includegraphics[width=0.9\columnwidth]{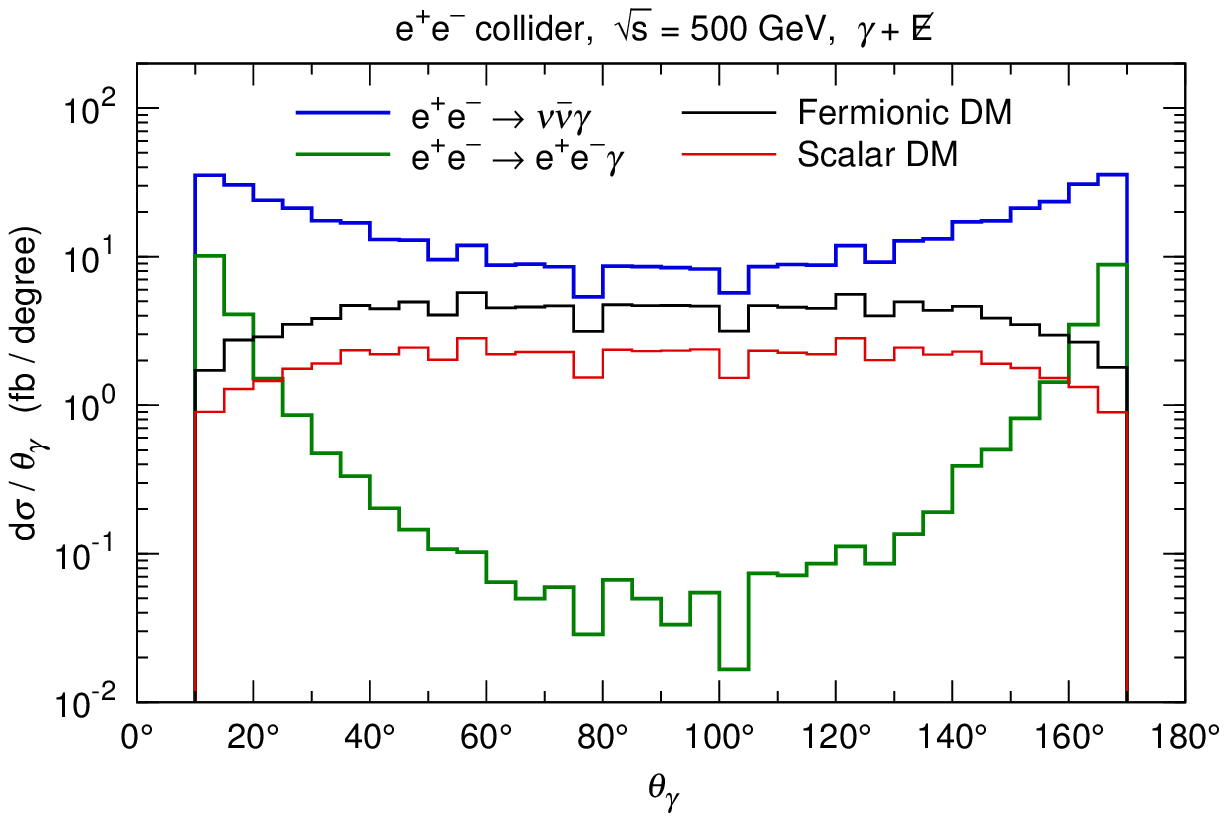}}
\subfigure[]{\includegraphics[width=0.9\columnwidth]{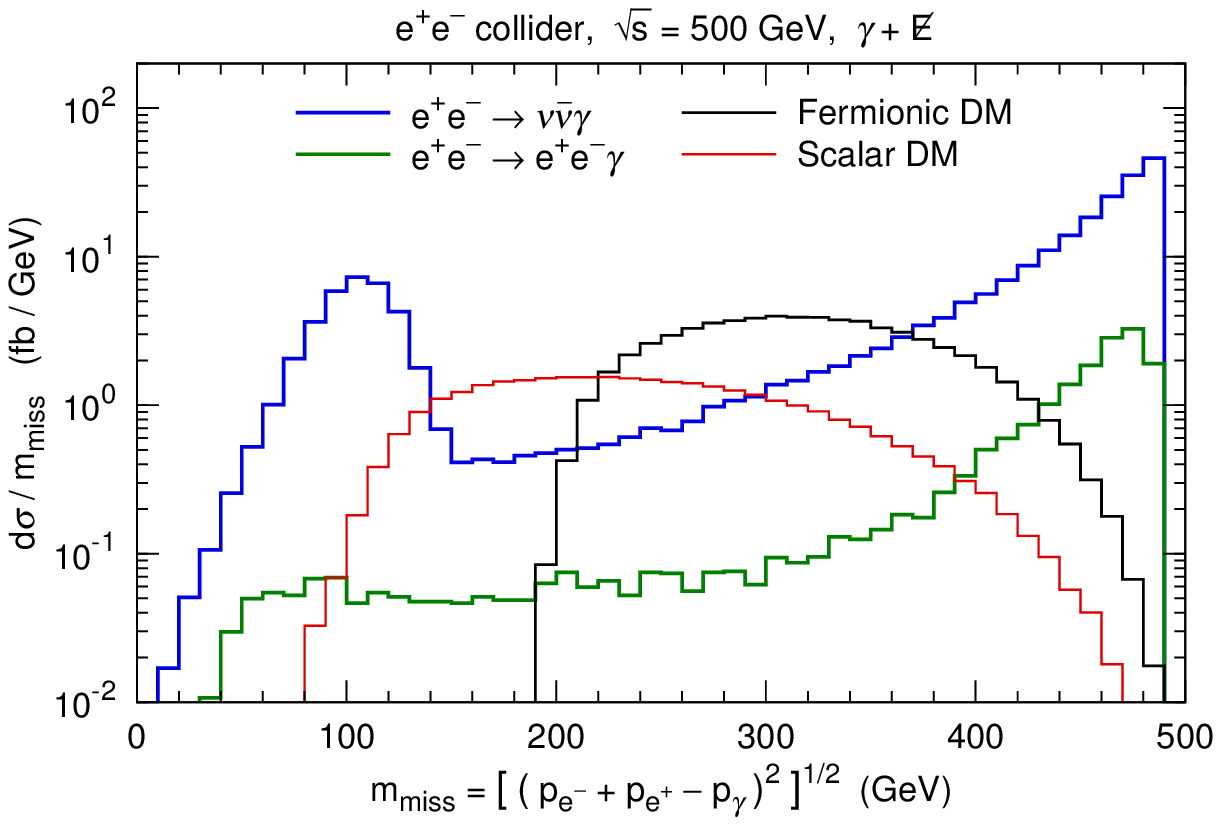}}
\caption{Distributions of the selected events
after Cut 1 for $\sqrt{s}=500\,\GeV$ are displayed.
For the fermionic (scalar) DM events, we use the following parameters
$\Lambda=200\,\GeV$ and $m_\chi=100\,(50)\,\GeV$.}
\label{fig:step1_dist}
\end{figure*}

\textbf{Cut 1:}
\textit{select the events containing a photon with $E_\gamma>10\,\GeV$
and $10^\circ<\theta_\gamma<170^\circ$.
Veto the events containing any other particle
with $E>10\,\GeV$ and $10^\circ<\theta<170^\circ$,
or with $E>50\,\GeV$ and $3^\circ\leq\theta\leq 177^\circ$.}

This cut picks up the so-called $\gamma+\missE$ events. In order to reliably identify a photon, the monophoton
is required to be energetic enough ($E_\gamma>10\,\GeV$) and not close to the beam axis ($10^\circ<\theta_\gamma<170^\circ$). The veto on other particles would effectively eliminate the events from $e^+e^-\to e^+e^-\gamma$ and other background events.

Using the case of $\sqrt{s}=500\,\GeV$ as an example,
we plot the $E_\gamma$, $\pT^\gamma$, $\theta_\gamma$,
and $m_\mathrm{miss}$ distributions of the selected events after Cut 1 in Fig.~\ref{fig:step1_dist}.
Here the missing mass $m_\mathrm{miss}$ is defined as
$m_\mathrm{miss} = \sqrt{(p_{e^-}+p_{e^+}-p_\gamma)^2}$,
where $p_{e^-}$ ($p_{e^+}$) is the 4-momentum
of the initial electron (positron).
For the fermionic (scalar) DM events, the distributions shown in the figure correspond to $\Lambda=200\,\GeV$ and $m_\chi=100\,(50)\,\GeV$.

\textbf{Cut 2:}
\textit{for $\sqrt{s}=250\,(500)\,\GeV$, veto the events with
$70\,(50)\,\GeV<m_\mathrm{miss}<110\,(130)\,\GeV$.
For $\sqrt{s}=1\,(3)\,\TeV$, veto the events with
$m_\mathrm{miss}<200\,(500)\,\GeV$.}

As clearly shown in Fig.~\ref{fig:step1_dist},
the $E_\gamma$ and $m_\mathrm{miss}$ distributions
of the background $e^+e^-\to\nu\bar\nu\gamma$ have peaks at
$\sim 240\,\GeV$ and at $\sim 100\,\GeV$ for $\sqrt{s}=500\,\GeV$,
respectively. These two peaks are correlated.
For the events coming from the $2\to2$ process $e^+e^-\to Z\gamma$,
$E_\gamma$ is determined to be $(s-m_Z^2)/(2\sqrt{s})$ and
the constructed $m_\mathrm{miss}$ equals to the $Z$ boson mass.
Therefore, this cut would eliminate
the backgrounds from $e^+e^-\to Z(\to\nu\bar\nu)\gamma$.
As $\sqrt{s}$ increases, the width of the peak
in the $m_\mathrm{miss}$ distribution expands
due to the imprecise measurements of the photon energy and momentum, and we extend the rejecting range of $m_\mathrm{miss}$ to take it into account.

If the $Z$ boson is produced in the $s$-channel, the cross section of $e^+e^-\to Z(\to\nu\bar\nu)\gamma$ decreases quickly as $\sqrt{s}$ increases. For $\sqrt{s} = 1\,\TeV$ and $3\,\TeV$, the event number from $e^+e^-\to Z(\to\nu\bar\nu)\gamma$ is quite small and can be negligible. However, when we suppress the background from $W$ boson exchanged Feynman diagrams by beam polarization in Sec.~\ref{sec:pol}, this cut can still be useful.

\textbf{Cut 3:}
\textit{requires the photon with $30^\circ<\theta_\gamma<150^\circ$.}

\textbf{Cut 4:}
\textit{requires the photon with $\pT^\gamma>\sqrt{s}/10$.}

For the 3-body production processes $e^+e^-\to\nu\bar\nu\gamma$ and $e^+e^-\to e^+e^-\gamma$, the photons come from initial state radiation and tend to be soft and collinear, as shown in the $E_\gamma$, $\pT^\gamma$, and $\theta_\gamma$ distributions in Fig.~\ref{fig:step1_dist}. On the other hand, for the signals, the photons are more energetic and their $\theta_\gamma$ distributions are rather flat. Consequently, Cuts 3 and 4 eliminate most of the $e^+e^-\to\nu\bar\nu\gamma$ events and remove almost all the $e^+e^-\to e^+e^-\gamma$ events without losing too much signal events.

\begin{table}[!htbp]
\caption{Cross sections $\sigma$ of all processes and signal significances $S/\sqrt{B}$
after each cut at $\sqrt{s}=500\,\GeV$ are tabulated.
For the fermionic (scalar) DM, we use $\Lambda=200\,\GeV$ and $m_\chi=100\,(50)\,\GeV$ as input.
The significances are computed by assuming the dataset with an integrated luminosity of $1\,\ifb$.}
\label{tab:cuts}
\begin{center}
\renewcommand{\arraystretch}{1.2}
\setlength\tabcolsep{0.5em}
\begin{tabular}{ccccccc}
\hline\hline
 & $\nu\bar\nu\gamma$ & $e^+e^-\gamma$
 & \multicolumn{2}{c}{Fermionic DM} & \multicolumn{2}{c}{Scalar DM} \\
 & $\sigma$\,($\fb$) & $\sigma$\,($\fb$)
 & $\sigma$\,($\fb$) & $S/\sqrt{B}$
 & $\sigma$\,($\fb$)& $S/\sqrt{B}$ \\
\hline
Cut 1 & 2415.2 & 173.0 & 646.8 & 12.7 & 321.4 & 6.3 \\
Cut 2 & 2102.5 & 168.6 & 646.8 & 13.6 & 308.2 & 6.5 \\
Cut 3 & 1161.1 & 16.8  & 538.0 & 15.7 & 255.9 & 7.5 \\
Cut 4 & 254.5  & 1.9   & 520.7 & 32.5 & 253.9 & 15.8 \\
\hline\hline
\end{tabular}
\end{center}
\end{table}

\begin{figure*}[!htbp]
\centering
\subfigure[]{\includegraphics[width=0.9\columnwidth]{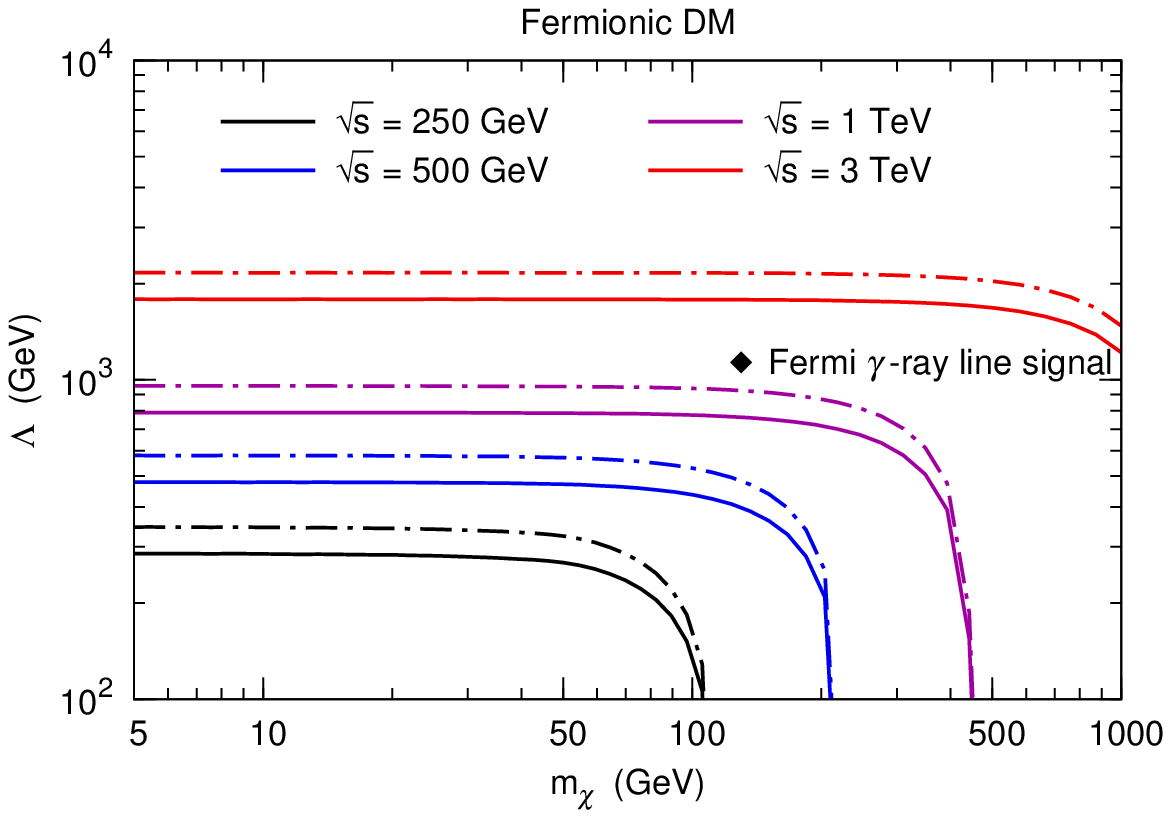}}
\subfigure[]{\includegraphics[width=0.9\columnwidth]{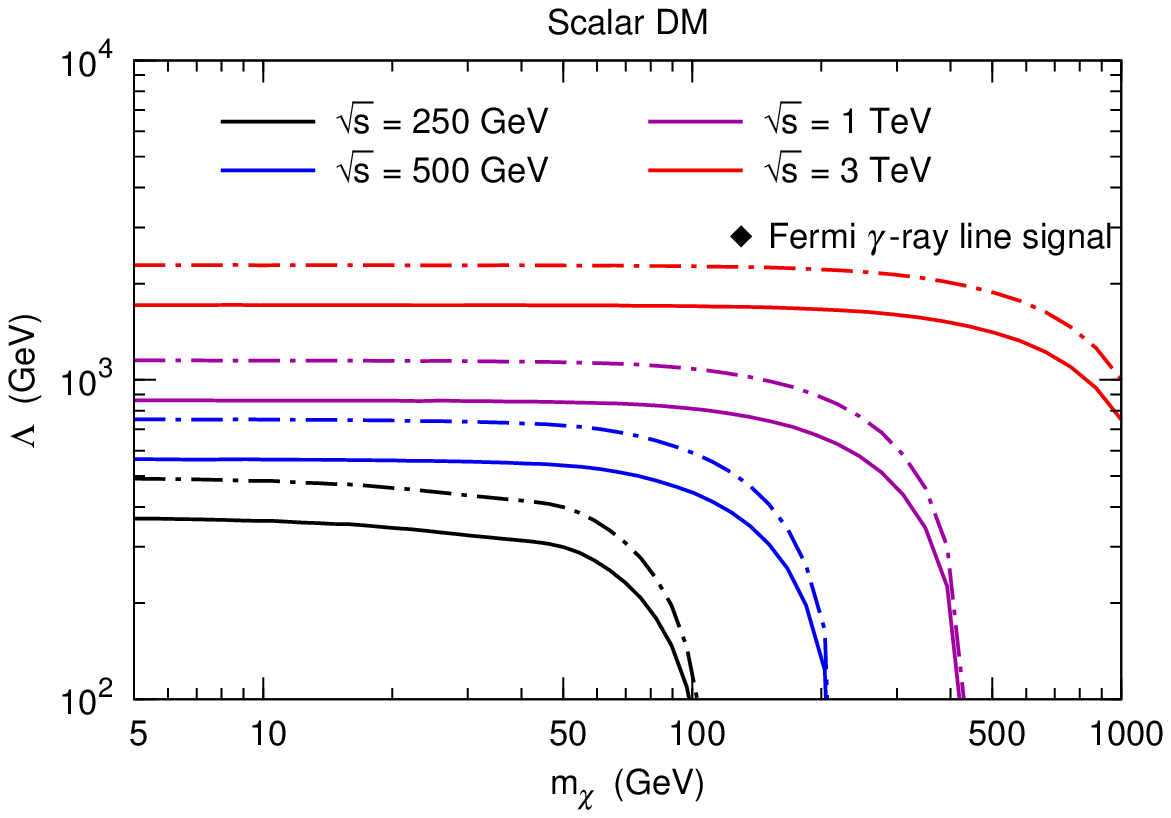}}
\caption{The $3\sigma$ reaches in the $m_\chi-\Lambda$ plane
at $e^+e^-$ colliders with $\sqrt{s}=250\,\GeV$, $500\,\GeV$, $1\,\TeV$,
and $3\,\TeV$ are shown.
The left (right) plot is for the fermionic (scalar) DM with integrated luminosities of
$100\,\ifb$ (solid lines) and $1000\,\ifb$ (dot-dashed lines), respectively.
The black diamonds denote the parameter points for explaining the Fermi $\gamma$-ray line signal.}
\label{fig:reaches}
\end{figure*}

The SM backgrounds are highly suppressed after imposing all the cuts.
As an illustration, in Table~\ref{tab:cuts}
we tabulate the cross sections of the backgrounds and the signals after each cut at an $e^+e^-$ collider with $\sqrt{s}=500\,\GeV$. We can see that the $e^+e^-\to\nu\bar\nu\gamma$ background is reduced by almost an order of magnitude, and only one percent of the $e^+e^-\to e^+e^-\gamma$ background survives.

We define $S/\sqrt{B}$ as the signal significance, where $S$ is
the number of signal events and $B$ is the total number of background events.
Table~\ref{tab:cuts} also lists the signal significances for
the fermionic (scalar) DM with $\Lambda=200\,\GeV$ and $m_\chi=100\,(50)\,\GeV$,
by assuming a dataset with an integrated luminosity of $1\,\ifb$.
The $3\sigma$ reaches in the $m_\chi-\Lambda$ plane
are shown in Fig.~\ref{fig:reaches}.

As mentioned above, in order to interpret the Fermi $\gamma$-ray line signal, the required $\Lambda$ of the fermionic DM are smaller than that of the scalar DM. At a $3\,\TeV$ $e^+e^-$ collider, we find that the parameter point corresponding to the Fermi signal for the fermionic DM can be easily confirmed or rejected.
On the other hand, even with a data set of $1000\,\ifb$, the parameter point for the scalar DM might be challenging.

\begin{figure*}[!htbp]
\centering
\subfigure[]{\includegraphics[width=0.9\columnwidth]{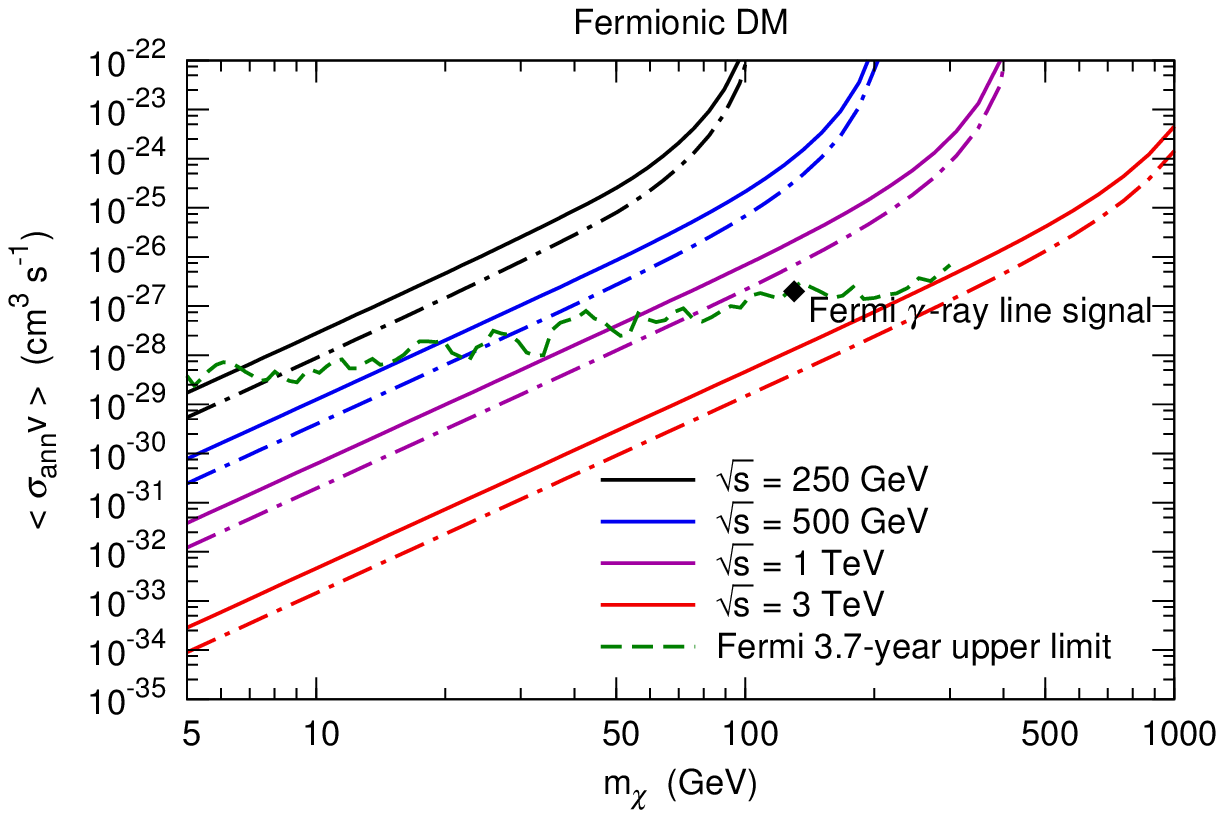}}
\subfigure[]{\includegraphics[width=0.9\columnwidth]{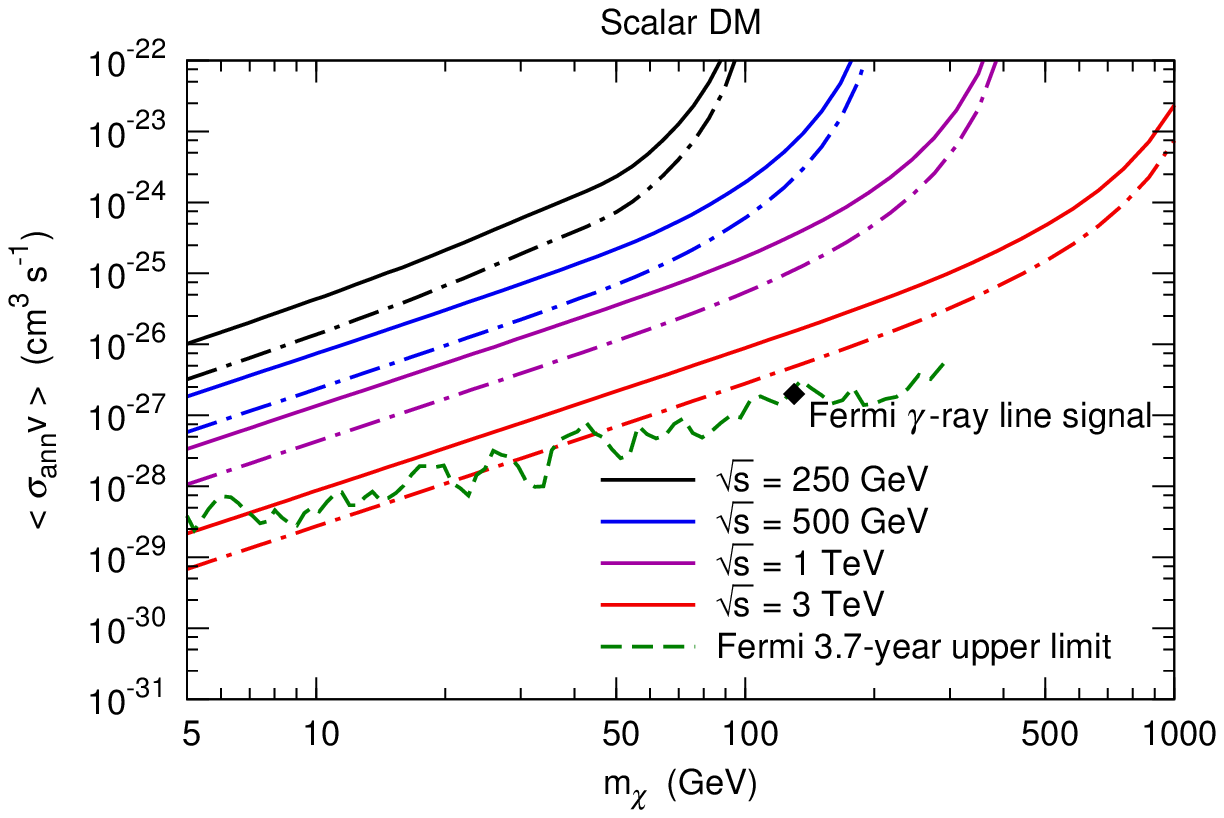}}
\caption{The $3\sigma$ reaches in the
$m_\chi-\left<\sigma_\mathrm{ann}v\right>$ plane for the fermionic DM (left)
and for the scalar DM (right) at $e^+e^-$ colliders are shown.
The lines denote the datasets of $100\,\ifb$ (solid lines) and $1000\,\ifb$
(dot-dashed lines), respectively.
The black diamonds denote the Fermi $\gamma$-ray line signal.
The green dashed lines indicate the Fermi 3.7-year 95\% CL upper limit
on DM annihilation cross section into $\gamma\gamma$.}
\label{fig:sv}
\end{figure*}

We convert the $3\sigma$ reaches from the $m_\chi-\Lambda$ plane to the
$m_\chi - \left<\sigma_\mathrm{ann}v\right>$ plane,
as shown in Fig.~\ref{fig:sv}. Using 3.7-year Fermi-LAT data,
the Fermi-LAT collaboration set some 95\% CL upper limits
on DM annihilation cross section into $\gamma\gamma$ \cite{Fermi-LAT:2013uma}.
For comparison, the limit from the region of R41~ given by Fermi-LAT collaboration is also shown in Fig.~\ref{fig:sv}. Note that the upper limits given in Ref.~\cite{Fermi-LAT:2013uma} corresponds to the case that the DM particle and its antiparticle are identical. Since we consider DM particles as Dirac fermions or complex scalar particles in this work, the DM particle is different to its antiparticle. In order to compensate this difference, the limit plotted in Fig.~\ref{fig:sv} has been scaled up by a factor of 2
(a similar treatment can be found in Ref.~\cite{ATLAS:2012ky}).

With a $100\,\ifb$ dataset we find that the fermionic DM searching at $e^+e^-$ colliders
could explore deeper than Fermi-LAT for light DM particles,
and the $3\sigma$ reach at a $3\,\TeV$ collider would be lower than the Fermi upper limit for $5\leq m_\chi\leq 300\,\GeV$.
However, the scalar DM searching at $e^+e^-$ colliders
would be challenging and would need a $\mathcal{O}(10^3)\,\ifb$ dataset  to make
the collider reaches comparable to the Fermi-LAT upper limit.

\section{Beam polarization \label{sec:pol}}

Polarized beams will be available at future $e^+e^-$ colliders.
Since the electroweak part of the SM is chiral,
appropriate beam polarization may be helpful to reduce SM backgrounds and to increase new physics signals~\cite{MoortgatPick:2005cw}. In Ref.~\cite{Dreiner:2006sb}, it was demonstrated that polarized beams can significantly suppress the background $e^+e^-\to\nu\bar\nu\gamma$.

For a process at an $e^+e^-$ collider with polarized beams, the cross section can be expressed as~\cite{MoortgatPick:2005cw}
\begin{eqnarray}
\sigma(P_{e^-},P_{e^+})
&=& \frac{1}{4} \big[
 (1+P_{e^-})(1+P_{e^+}) \sigma_\mathrm{RR}
\nonumber\\
&&~ +(1-P_{e^-})(1-P_{e^+}) \sigma_\mathrm{LL}
\nonumber\\
&&~ +(1+P_{e^-})(1-P_{e^+}) \sigma_\mathrm{RL}
\nonumber\\
&&~ +(1-P_{e^-})(1+P_{e^+}) \sigma_\mathrm{LR}
\big],
\end{eqnarray}
where $P_{e^\pm}$ is the longitudinal degree of $e^\pm$ beam polarization.
$P_{e^\pm}>0$ ($P_{e^\pm}<0$) corresponds to the right-handed (left-handed)
polarization. $\sigma_\mathrm{RL}$ denotes the cross section
for the completely right-handed polarized $e^-$ beam ($P_{e^-}=+1$)
and the completely left-handed polarized $e^+$ beam ($P_{e^+}=-1$).
The definitions of $\sigma_\mathrm{LR}$, $\sigma_\mathrm{RR}$,
and $\sigma_\mathrm{LL}$ are analogous.
In Fig.~\ref{fig:polar_Xsec}, we show the polarized cross sections
for the dominant SM background $e^+e^-\to\nu\bar\nu\gamma$, the fermionic DM production $e^+e^-\to\chi\bar\chi\gamma$,
and the scalar DM production $e^+e^-\to\chi\chi^*\gamma$
at an $e^+e^-$ collider with $\sqrt{s}=500\,\GeV$,
after applying the kinematic cuts $E_\gamma>10\,\GeV$ and
$10^\circ<\theta_\gamma<170^\circ$.

\begin{figure*}[!htbp]
\centering
\subfigure[]{\includegraphics[width=0.65\columnwidth]{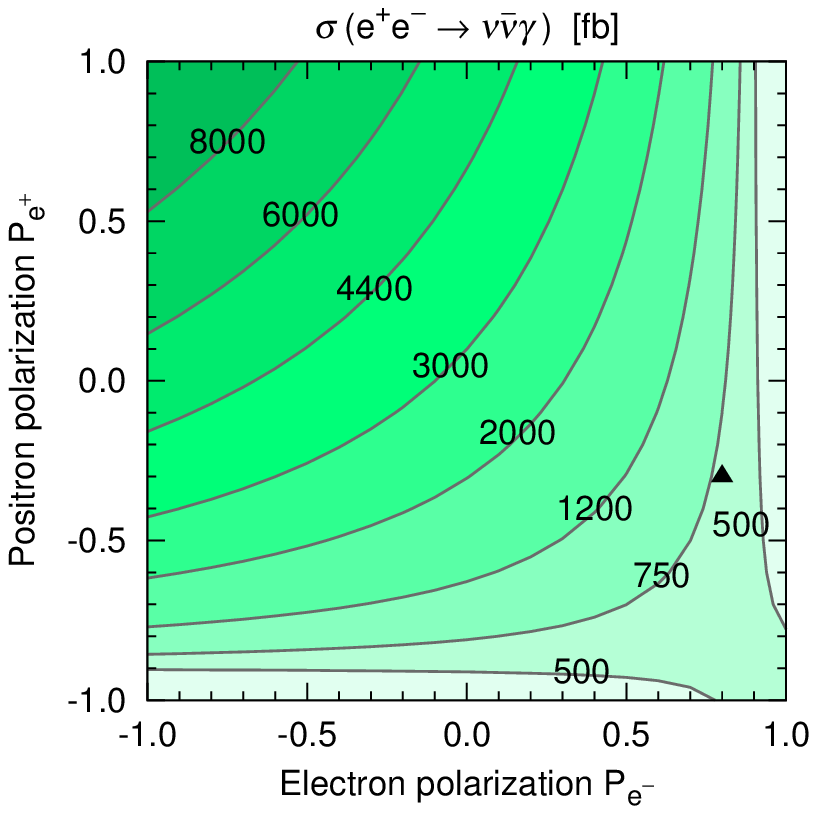}}
\subfigure[]{\includegraphics[width=0.65\columnwidth]{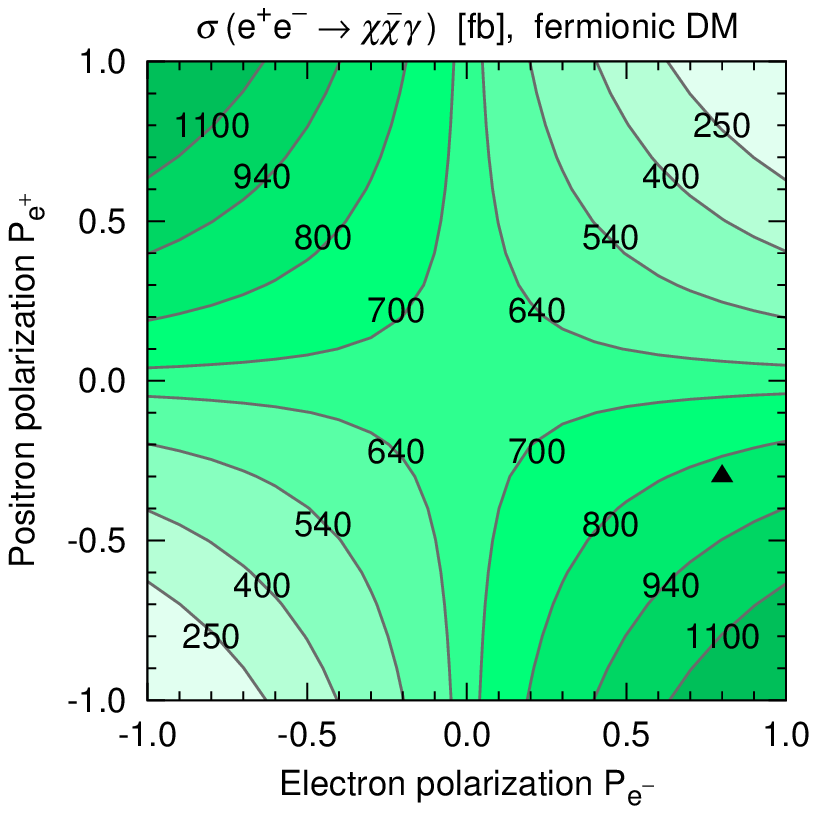}}
\subfigure[]{\includegraphics[width=0.65\columnwidth]{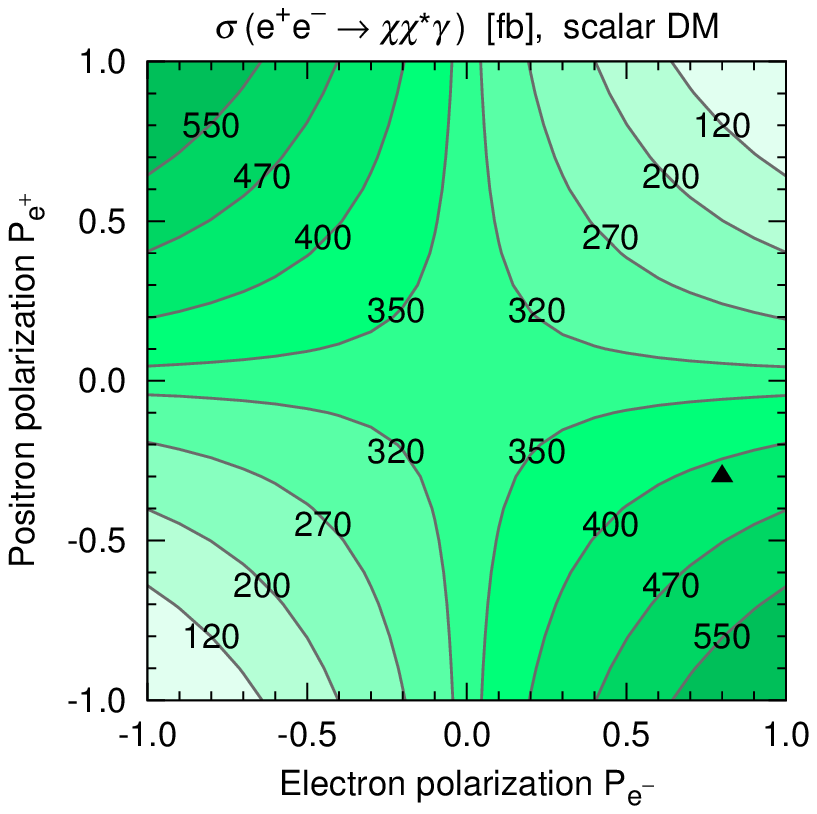}}
\caption{Contour lines of the polarized cross sections for the dominant SM background
$e^+e^-\to\nu\bar\nu\gamma$ (left), the fermionic DM production
$e^+e^-\to\chi\bar\chi\gamma$ (center), and the scalar DM production
$e^+e^-\to\chi\chi^*\gamma$ (right) are shown by using
an $e^+e^-$ collider with $\sqrt{s}=500\,\GeV$ as a show case.
The kinematic cuts $E_\gamma>10\,\GeV$ and
$10^\circ<\theta_\gamma<170^\circ$ are applied.
For the fermionic (scalar) DM, the parameters are adopted to be
$\Lambda=200\,\GeV$ and $m_\chi=100\,(50)\,\GeV$.
The triangles denote the realistic polarization configuration of
$(P_{e^-},P_{e^+})=(0.8,-0.3)$.}
\label{fig:polar_Xsec}
\end{figure*}

For $e^+e^-\to\nu\bar\nu\gamma$, the Feynman diagrams involving $t$-channel
$W$ boson exchange cannot contribute to $\sigma_\mathrm{RL}$,
$\sigma_\mathrm{RR}$, and $\sigma_\mathrm{LL}$, since the $W$ boson couples to
neither right-handed $e^-$ nor left-handed $e^+$.
On the other hand, the $Z$ boson does not couple to
$e^-_\mathrm{R}e^+_\mathrm{R}$ or $e^-_\mathrm{L}e^+_\mathrm{L}$,
while the coupling to $e^-_\mathrm{L} e^+_\mathrm{R}$
is stronger than that to $e^-_\mathrm{R} e^+_\mathrm{L}$.
Thus the Feynman diagrams involving $s$-channel
$Z$ boson exchange have more contributions to $\sigma_\mathrm{LR}$
than to $\sigma_\mathrm{RL}$.
Consequently, the cross section for $e^+e^-\to\nu\bar\nu\gamma$
vanishes with $(P_{e^-},P_{e^+})=(+1,+1)$ or with $(P_{e^-},P_{e^+})=(-1,-1)$,
while $\sigma_\mathrm{LR} (e^+e^-\to\nu\bar\nu\gamma)$ is larger than
$\sigma_\mathrm{RL} (e^+e^-\to\nu\bar\nu\gamma)$ by a factor of $\sim 20$
at $\sqrt{s}=500\,\GeV$.

For the DM production process $e^+e^-\to\chi\bar\chi\gamma$
or $e^+e^-\to\chi\chi^*\gamma$, $\sigma_\mathrm{RR}$ and $\sigma_\mathrm{LL}$
vanish because the corresponding processes cannot preserve angular momentum.
The angular momentum quantum numbers
of the $e^-_\mathrm{R}e^+_\mathrm{R}$ system
and the $e^-_\mathrm{L}e^+_\mathrm{L}$ system are both 0,
while the exchanged $s$-channel photon has a spin of 1.
On the other hand, $\sigma_\mathrm{LR}$ is equal to $\sigma_\mathrm{RL}$,
and the unpolarized cross section is just a half of
either $\sigma_\mathrm{LR}$ or $\sigma_\mathrm{RL}$.

\begin{figure*}[!htbp]
\centering
\subfigure[]{\includegraphics[width=0.9\columnwidth]{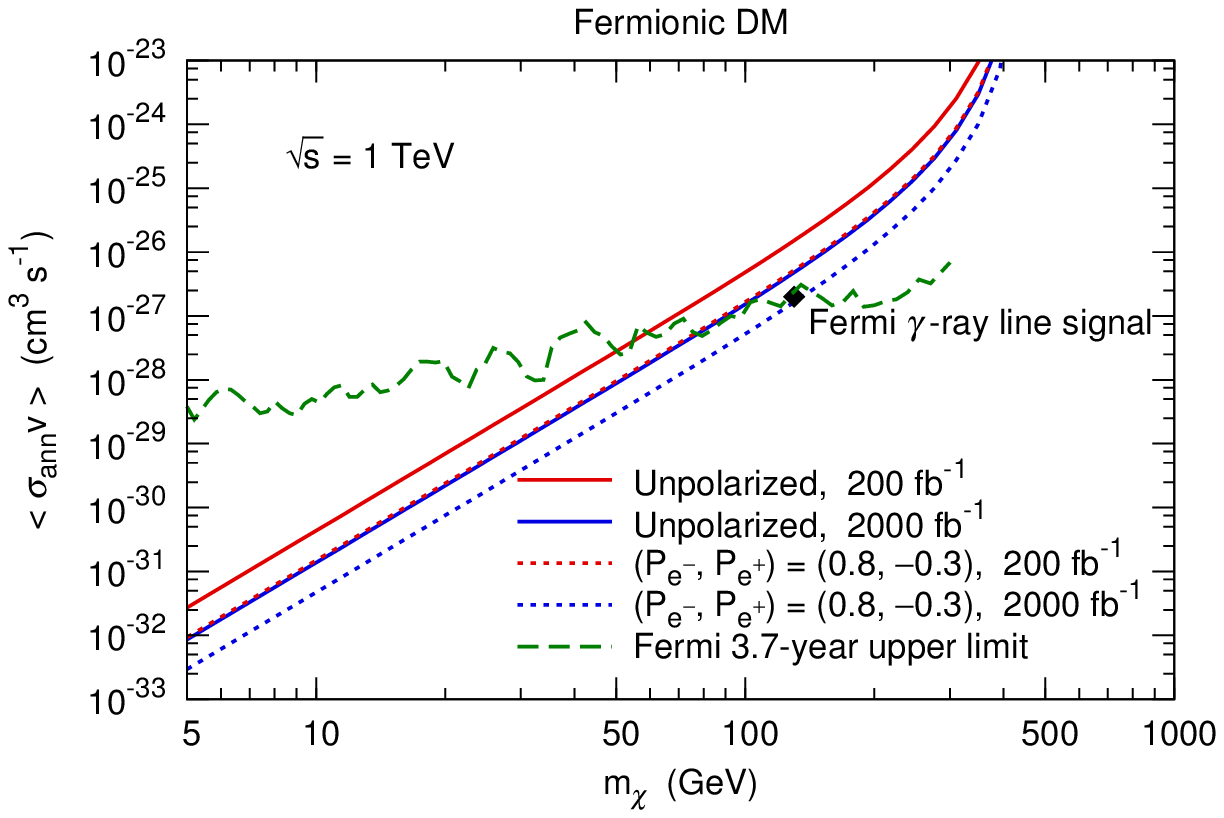}}
\subfigure[]{\includegraphics[width=0.9\columnwidth]{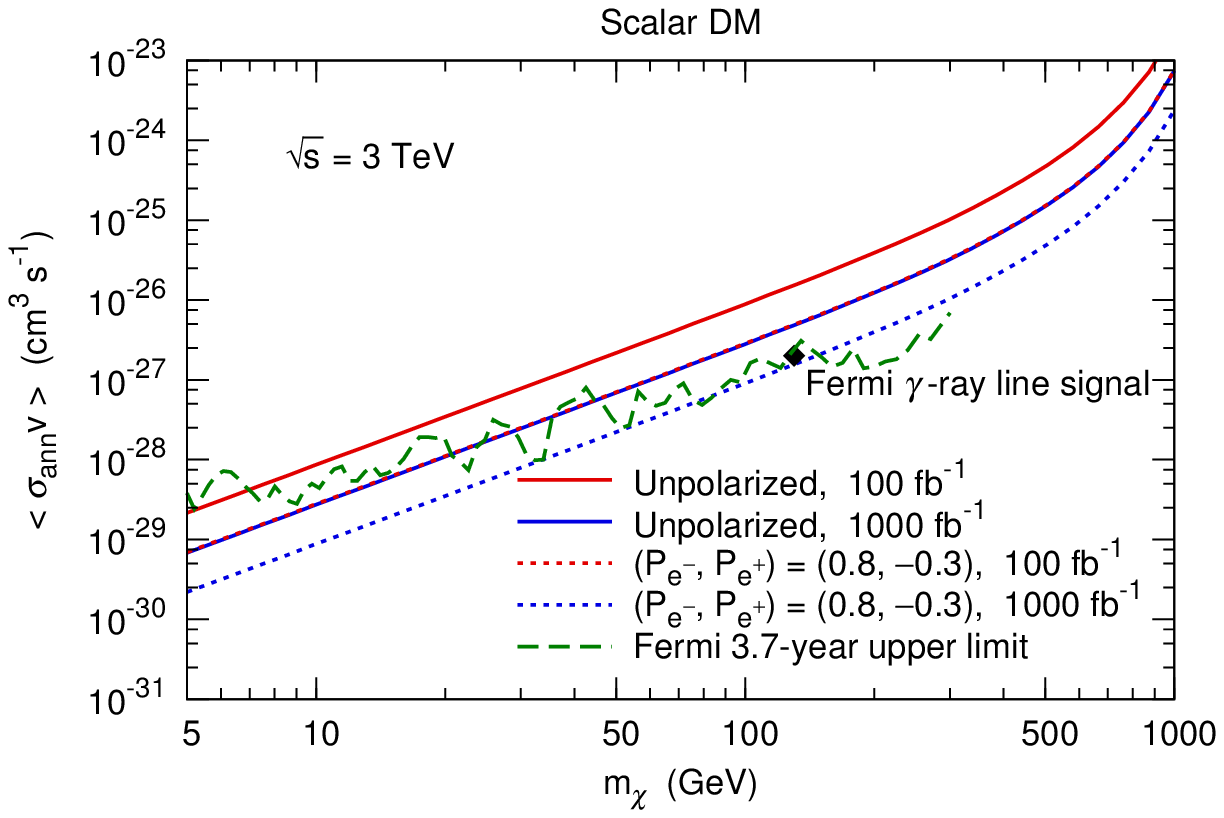}}
\caption{The $3\sigma$ reaches in the $m_\chi-\left<\sigma_\mathrm{ann}v\right>$ plane
at $e^+e^-$ colliders for unpolarized beams (solid lines) and
for the polarized beams with $(P_{e^-},P_{e^+})=(0.8,-0.3)$ (dot lines) are demonstrated.
For the fermionic DM (left), the collision energy is assumed to be $\sqrt{s}=1\,\TeV$,
while for the scalar DM (right) $\sqrt{s}=3\,\TeV$.}
\label{fig:polar_sv}
\end{figure*}

From the analysis above, we can see that the maximal signal significances
can be achieved with the beam polarization of $(P_{e^-},P_{e^+})=(+1,-1)$.
However, a beam with a polarization degree of 100\% can not be realized
in practice. Instead, beams with a polarization degree of 80\% for $e^-$ and a polarization
degree of 30\% for $e^+$ are realistic at the ILC~\cite{Behnke:2013xla}.
Although higher degree of beam polarization could be available in the future, we conservatively consider a polarization configuration of $(P_{e^-},P_{e^+})=(0.8,-0.3)$.
Then at $\sqrt{s}=500\,\GeV$,
the polarized cross section for $e^+e^-\to\nu\bar\nu\gamma$
is just $\sim 25\%$ of the unpolarized cross section,
while both the cross sections of $e^+e^-\to\chi\bar\chi\gamma$
and $e^+e^-\to\chi\chi^*\gamma$ are enhanced by a factor of 1.24. Therefore, the choice of polarization configurations are powerful to suppress the background process $e^+e^-\to\nu\bar\nu\gamma$ and to enhance the production rate of DM signals, simultaneously.

We apply the same event cuts used in Sec.~\ref{sec:reach} and give the $3\sigma$ reaches at future $e^+e^-$ colliders. In Fig.~\ref{fig:polar_sv}, we show the $3\sigma$ reaches in the $m_\chi-\left<\sigma_\mathrm{ann}v\right>$ plane for the polarized beams.
For comparison, the results for unpolarized beams are also plotted. We find that using the polarized beams is roughly equivalent to increasing the integrated luminosity by an order of magnitude. After using the polarized beams, for the fermionic (scalar) DM , a $2000\,\ifb$ ($1000\,\ifb$) dataset would be just sufficient to test the Fermi $\gamma$-ray line signal
at an $e^+e^-$ collider with $\sqrt{s}=1\,\TeV$ ($3\,\TeV$).

\section{Unitarity bounds \label{sec:uni}}

When the collision energy is sufficiently high compared with the typical energy scale of the effective operator, the effective operator description would fail. If this occurs, the $3\sigma$ reaches derived above would be invalid.

In scattering theories, the unitarity of the $S$-matrix
corresponds to the conservation of probability.
In order to preserve probability, the unitarity at any order
of a perturbative theory should not be violated.
When a process described by an effective theory violate the unitarity, it means that the theory should not be used for this process and a UV-complete theory should be introduced for the right description. It has been noticed that the effective operator treatment for DM searches at the LHC should be carefully checked by verifying the $S$-matrix unitarity~\cite{Shoemaker:2011vi,Fox:2012ee}.\footnote{Another way
to discuss the validity of effective theory is to check
if the couplings stay in the perturbative regime~\cite{Busoni:2013lha}.}
Thus we should consider this verification for the processes we studied at $e^+e^-$ colliders.

Since the leading process of DM production is
the $2\to3$ process $e^+e^-\to\chi\chi\gamma$,
the traditional unitarity condition for $2\to2$ processes can not be directly applied here. Thus we need to have a general unitarity condition for $2 \to n$ processes in order to study the unitarity bound for the process $e^+e^-\to\chi\chi\gamma$.
For this purpose, firstly, we recapitulate the derivation of
the traditional unitarity condition for $2 \to 2$ processes according to the standard method presented in Ref.~\cite{Bjorken:1965zz}.

Inserting $S=1+iT$ into the unitarity condition of the $S$-matrix $S^\dag S=1$, we have
\begin{equation}
-i(T-T^\dag)=T^\dag T.
\label{eq:T}
\end{equation}
Using the identities $\left<\beta\right|T\left|\alpha\right>=(2\pi)^4
\delta^{(4)}(p_\alpha-p_\beta)\mathcal{M}_{\alpha\to\beta}$ and
$\left<\beta\right|T^\dag\left|\alpha\right>=
\left<\alpha\right|T\left|\beta\right>^*$,
we can express the matrix element of the left-hand side of Eq.~\eqref{eq:T} between the initial state $\left|\alpha\right>$ and
the final state $\left|\beta\right>$ as
\begin{eqnarray}
-i\left<\beta\right| &\,& T-T^\dag\left|\alpha\right>=
\nonumber\\
&&-i(2\pi)^4 \delta^{(4)}(p_\alpha-p_\beta)
(\mathcal{M}_{\alpha\to\beta}-\mathcal{M}_{\beta\to\alpha}^*).
\end{eqnarray}
For the right-hand side, we insert a complete set of
intermediate states ${\left|\gamma\right>}$:
\begin{eqnarray}
\left<\beta\right|T^\dag T\left|\alpha\right>
&=&\sum\limits_\gamma \int d\Pi_\gamma \left<\beta\right| T^\dag
\left|\gamma\right>\left<\gamma\right|T\left|\alpha\right>
\nonumber\\
&=& (2\pi)^4\delta^{(4)}(p_\alpha-p_\beta)
\sum\limits_\gamma \int d\Pi_\gamma
\mathcal{M}_{\beta\to\gamma}^*\mathcal{M}_{\alpha\to\gamma}
\nonumber\\
&&\times(2\pi)^4\delta^{(4)}(p_\alpha-p_\gamma),
\end{eqnarray}
where $d\Pi_\gamma \equiv \prod\limits_i\dfrac{d^3 p_{\gamma_i}}
{(2\pi)^3 2p_{\gamma_i}^0}$, denoting the phase space of intermediate states. Thus Eq.~\eqref{eq:T} becomes
\begin{eqnarray}
-i(\mathcal{M}_{\alpha\to\beta}-\mathcal{M}_{\beta\to\alpha}^*)
 &=& \sum\limits_\gamma \int d\Pi_\gamma
 \mathcal{M}_{\beta\to\gamma}^*\mathcal{M}_{\alpha\to\gamma}
\nonumber\\
&&\times
(2\pi)^4\delta^{(4)}(p_\alpha-p_\gamma).
\label{eq:uni_amp}
\end{eqnarray}

In the center-of-mass frame, the amplitude of a $2 \to 2$ process
$\mathcal{M}(s,\cos\theta)$ only depends on
$s$ and the scattering angle $\theta$.
We will suppress the dependence on $s$ in the following derivation.
For the elastic process $1+2 \to 1+2$
between Particle 1 and Particle 2 with masses $m_1$ and $m_2$,
we consider the following transitions of state:
\begin{eqnarray}
\alpha(p_1,p_2)&\to&\beta(q_1,q_2)\text{ with }
\mathcal{M}_\mathrm{el}(\cos\theta_{\alpha\beta}),
\nonumber\\
\alpha(p_1,p_2)&\to&\gamma_\mathrm{el}(k_1,k_2)\text{ with }
\mathcal{M}_\mathrm{el}(\cos\theta_{\alpha\gamma}),
\nonumber\\
\beta(q_1,q_2)&\to&\gamma_\mathrm{el}(k_1,k_2)\text{ with }
\mathcal{M}_\mathrm{el}(\cos\theta_{\beta\gamma}),
\label{eq:state_trans_el}
\end{eqnarray}
where $\cos\theta_{\alpha\beta}=\hat{\mathbf{p}}_1\cdot\hat{\mathbf{q}}_1$,
$\cos\theta_{\alpha\gamma}=\hat{\mathbf{p}}_1\cdot\hat{\mathbf{k}}_1$,
and $\cos\theta_{\beta\gamma}=\hat{\mathbf{q}}_1\cdot\hat{\mathbf{k}}_1$.
Since $\mathcal{M}_{\alpha\to\beta}=\mathcal{M^*}_{\beta\to\alpha}
=\mathcal{M}_\mathrm{el}(\cos\theta_{\alpha\beta})$,
The unitarity condition \eqref{eq:uni_amp} becomes
\begin{eqnarray}
&&2\operatorname{Im}\mathcal{M}_\mathrm{el}(\cos\theta_{\alpha\beta})
\nonumber\\
&=& \int d\Pi_{\gamma_\mathrm{el}} \mathcal{M}_{\beta\to\gamma_\mathrm{el}}^*
\mathcal{M}_{\alpha\to\gamma_\mathrm{el}}
(2\pi)^4 \delta^{(4)}(p_\alpha-p_{\gamma_\mathrm{el}})
\nonumber\\
 &&\quad + \text{ inelastic terms}
 \nonumber\\
 &\geq& \frac{\beta_1}{32\pi^2} \int d\Omega_{k_1}
 \mathcal{M}_\mathrm{el}^*(\cos\theta_{\beta\gamma})
 \mathcal{M}_\mathrm{el}(\cos\theta_{\alpha\gamma}),
\label{ieq:M_el}
\end{eqnarray}
where $\beta_1\equiv\sqrt{1-4m_1^2/s}$ and
$d\Omega_{k_1}=d\phi_{k_1} d\cos\theta_{\alpha\gamma}$.

The $2 \to 2$ amplitude $\mathcal{M}(\cos\theta)$
can be expanded as partial waves:
\begin{eqnarray}
\mathcal{M}(\cos\theta) &=& 16\pi\sum_j (2j+1) a_j P_j(\cos\theta),
\nonumber\\
a_j &=& \frac{1}{32\pi} \int_{-1}^1 d\cos\theta
P_j(\cos\theta)\mathcal{M}(\cos\theta),
\end{eqnarray}
where $P_j(x)$ are Legendre polynomials.
After multiplying the both sides of the inequality \eqref{ieq:M_el}
by $(64\pi)^{-1}P_j(\cos\theta_{\alpha\beta})$
and integrating over $\cos\theta_{\alpha\beta}$, we can obtain
\begin{eqnarray}
\operatorname{Im} a_j^\mathrm{el}
&\geq& \frac{\beta_1}{8\pi}
\sum_{k,l} (2k+1)(2l+1) a_k^{\mathrm{el}*} a_l^\mathrm{el}
\int d\cos\theta_{\alpha\beta} d\Omega_{k_1}
\nonumber\\
&&\times P_j(\cos\theta_{\alpha\beta})
 P_k(\cos\theta_{\beta\gamma}) P_l(\cos\theta_{\alpha\gamma}).
\end{eqnarray}
Using the addition theorem for Legendre polynomials
(see e.g. Ref.~\cite{Arfken:math})
\begin{eqnarray}
P_k && (\cos\theta_{\beta\gamma})
= P_k(\cos\theta_{\alpha\beta}) {P_k}(\cos\theta_{\alpha\gamma})
\nonumber\\
&& + 2\sum_{m=1}^l \frac{(l-m)!}{(l+m)!} P_k^m(\cos\theta_{\alpha\beta})
P_k^m(\cos\theta_{\alpha\gamma}) \cos m\phi_{k_1}
\nonumber\\
\end{eqnarray}
and the orthogonality relation
\begin{equation}
\int_{-1}^1 d\cos\theta P_j(\cos\theta) P_k(\cos\theta)
=\frac{2}{2j+1} \delta_{jk},
\end{equation}
and carrying out all the integrations, we have
\begin{equation}
\operatorname{Im} a_j^\mathrm{el} \geq \beta_1 |a_j^\mathrm{el}|^2,
\label{ieq:Im_a_el}
\end{equation}
which is equivalent to
\begin{equation}
(\operatorname{Re}a_j^\mathrm{el})^2 + \left(\operatorname{Im}
a_j^\mathrm{el} - \frac{1}{2\beta_1} \right)^2 \leq \frac{1}{(2\beta_1)^2}.
\end{equation}
For the scattering of massless particles, $\beta_1=1$, and it implies
\begin{equation}
\left|\operatorname{Re}a_j^\mathrm{el}(s)\right|\leq\frac{1}{2},
\quad \forall j.
\end{equation}
This is the well-known unitarity condition for $2 \to 2$ elastic scattering.
It means that the real part of every amplitude partial wave cannot exceeds $1/2$.

Now we consider an inelastic process $1+2 \to 3+4$ by the transitions of state
\begin{eqnarray}
\alpha(p_1,p_2)&\to&\gamma_\mathrm{inel}(k_3,k_4)\text{ with }
\mathcal{M}_\mathrm{inel}(\cos\theta'_{\alpha\gamma}),
\nonumber\\
\beta(q_1,q_2)&\to&\gamma_\mathrm{inel}(k_3,k_4)\text{ with }
\mathcal{M}_\mathrm{inel}(\cos\theta'_{\beta\gamma}),
\end{eqnarray}
where $\cos\theta'_{\alpha\gamma}=\hat{\mathbf{p}}_1\cdot\hat{\mathbf{k}}_3$
and $\cos\theta'_{\beta\gamma}=\hat{\mathbf{q}}_1\cdot\hat{\mathbf{k}}_3$.
The masses of Particle 3 and Particle 4 are $m_3$ and $m_4$.
The initial state $\alpha(p_1,p_2)$ differs from another initial state $\beta(q_1,q_2)$
by an angle $\theta_{\alpha\beta}$, as in the case of Eq.~\eqref{eq:state_trans_el}.
We can re-express the inequality \eqref{ieq:M_el} by extracting
the term corresponding to $1+2 \to 3+4$ from the ``inelastic terms'':
\begin{eqnarray}
&&2\operatorname{Im}\mathcal{M}_\mathrm{el}(\cos\theta_{\alpha\beta})
\nonumber\\
&=& \int d\Pi_{\gamma_\mathrm{el}} \mathcal{M}_{\beta\to\gamma_\mathrm{el}}^*
\mathcal{M}_{\alpha\to\gamma_\mathrm{el}}
(2\pi)^4 \delta^{(4)}(p_\alpha-p_{\gamma_\mathrm{el}})
\nonumber\\
&& + \int d\Pi_{\gamma_\mathrm{inel}} \mathcal{M}_{\beta\to\gamma_\mathrm{inel}}^*
\mathcal{M}_{\alpha\to\gamma_\mathrm{inel}}
(2\pi)^4 \delta^{(4)}(p_\alpha-p_{\gamma_\mathrm{inel}})
\nonumber\\
 &&\quad + \text{ other inelastic terms}
\nonumber\\
 &\geq& \frac{\beta_1}{32\pi^2} \int d\Omega_{k_1}
 \mathcal{M}_\mathrm{el}^*(\cos\theta_{\beta\gamma})
 \mathcal{M}_\mathrm{el}(\cos\theta_{\alpha\gamma})
\nonumber\\
&& + \frac{\beta_3}{32\pi^2} \int d\Omega_{k_3}
 \mathcal{M}_\mathrm{inel}^*(\cos\theta'_{\beta\gamma})
 \mathcal{M}_\mathrm{inel}(\cos\theta'_{\alpha\gamma}),
\label{ieq:M_el_inel}
\end{eqnarray}
where $\beta_3\equiv\sqrt{1-4m_3^2/s}$ and
$d\Omega_{k_3}=d\phi_{k_3} d\cos\theta'_{\alpha\gamma}$.
As the derivation of \eqref{ieq:Im_a_el}, we have
\begin{equation}
\operatorname{Im} a_j^\mathrm{el} \geq \beta_1 |a_j^\mathrm{el}|^2
+ \beta_3 |a_j^\mathrm{inel}|^2,
\end{equation}
which is equivalent to
\begin{equation}
\frac{1}{4\beta_1} - \beta_1 \left[(\operatorname{Re}a_j^\mathrm{el})^2
+ \left(\operatorname{Im}a_j^\mathrm{el} - \frac{1}{2\beta_1}\right)^2\right]
\geq \beta_3 |a_j^\mathrm{inel}|^2.
\end{equation}
Thus for massless incoming particles,
\begin{equation}
\left|a_j^\mathrm{inel}(s)\right| \leq \frac{1}{2\sqrt{\beta_3}},
\quad \forall j.
\label{eq:a_inel_uni}
\end{equation}
This is the unitarity condition for $2 \to 2$ inelastic scattering
(see e.g. Ref.~\cite{Marciano:1989ns}).

In order to derive a general unitarity condition for
$2 \to n$ inelastic scattering, we consider the transitions of state
$\alpha(p_1,p_2)\to\gamma_n$ and $\beta(q_1,q_2)\to\gamma_n$,
where $\gamma_n$ denotes a state with $n$ particles.
As the derivation of \eqref{ieq:M_el_inel}, we can have
\begin{eqnarray}
&&2\operatorname{Im}\mathcal{M}_\mathrm{el}(\cos\theta_{\alpha\beta})
\nonumber\\
&=& \int d\Pi_{\gamma_\mathrm{el}} \mathcal{M}_{\beta\to\gamma_\mathrm{el}}^*
\mathcal{M}_{\alpha\to\gamma_\mathrm{el}}
(2\pi)^4 \delta^{(4)}(p_\alpha-p_{\gamma_\mathrm{el}})
\nonumber\\
&& + \int d\Pi_{\gamma_n} \mathcal{M}_{\beta\to\gamma_n}^*
\mathcal{M}_{\alpha\to\gamma_n}
(2\pi)^4 \delta^{(4)}(p_\alpha-p_{\gamma_n})
\nonumber\\
 &&\quad + \text{ other inelastic terms}
\nonumber\\
 &\geq& \frac{\beta_1}{32\pi^2} \int d\Omega_{k_1}
 \mathcal{M}_\mathrm{el}^*(\cos\theta_{\beta\gamma})
 \mathcal{M}_\mathrm{el}(\cos\theta_{\alpha\gamma})
\nonumber\\
&& + \int d\Pi_{\gamma_n} \mathcal{M}_{\beta\to\gamma_n}^*
\mathcal{M}_{\alpha\to\gamma_n}
(2\pi)^4 \delta^{(4)}(p_\alpha-p_{\gamma_n}).
\end{eqnarray}
Expressing the elastic terms in this inequality and using the partial wave expansion, we have
\begin{equation}
\operatorname{Im} a_j^\mathrm{el} \geq \beta_1 |a_j^\mathrm{el}|^2
+ |b_j^\mathrm{inel}|^2,
\label{ieq:Im_a_el_b}
\end{equation}
where
\begin{eqnarray}
|b_j^\mathrm{inel}|^2 &\equiv&
\frac{1}{64\pi} \int d\cos\theta_{\alpha\beta} P_j(\cos\theta_{\alpha\beta})
\nonumber\\
&&\quad\times \int d\Pi_{\gamma_n} \mathcal{M}_{\beta\to\gamma_n}^*
\mathcal{M}_{\alpha\to\gamma_n}
(2\pi)^4 \delta^{(4)}(p_\alpha-p_{\gamma_n})
\nonumber\\
\label{eq:b_inel_def}
\end{eqnarray}
is a new quantity to express the unitarity condition
for $2 \to n$ inelastic scattering.
According to \eqref{ieq:Im_a_el_b}, it is straightforward to have
\begin{eqnarray}
|b_j^\mathrm{inel}|^2 &\leq&
\frac{1}{4\beta_1} - \beta_1 \left[(\operatorname{Re}a_j^\mathrm{el})^2
+ \left(\operatorname{Im}a_j^\mathrm{el} - \frac{1}{2\beta_1}\right)^2\right]
\nonumber\\
&\leq& \frac{1}{4\beta_1}.
\end{eqnarray}
Thus, for massless incoming particles, a general unitarity condition
for $2 \to n$ inelastic scattering is obtained
\begin{equation}
|b_j^\mathrm{inel}(s)| \leq \frac{1}{2}, \quad\forall j.
\label{ieq:b_inel}
\end{equation}
For $2 \to 2$ inelastic scattering,
we can have $|b_j^\mathrm{inel}|^2 = \beta_3 |a_j^\mathrm{inel}|^2$,
and then Eq.~\eqref{ieq:b_inel} elegantly goes back to Eq.~\eqref{eq:a_inel_uni}.

\begin{figure*}[!htbp]
\centering
\subfigure[]{\includegraphics[width=0.75\columnwidth]{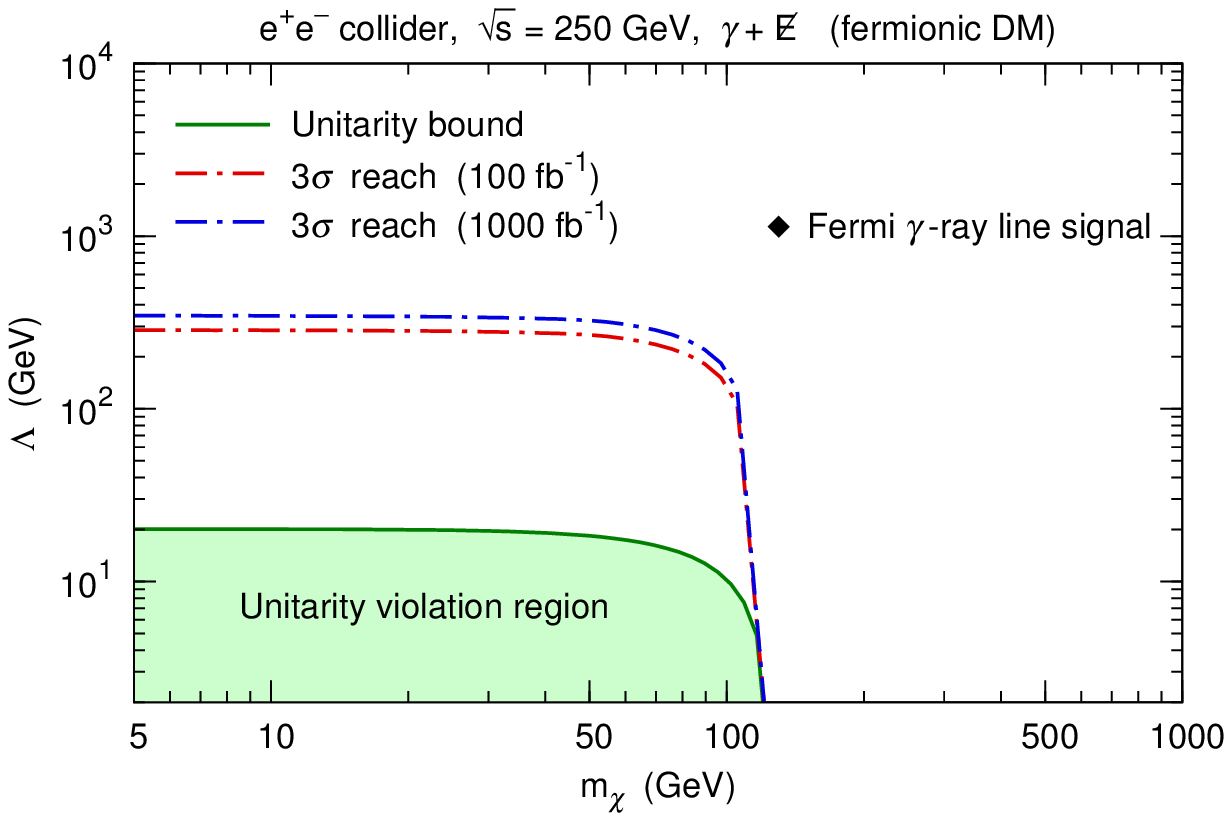}}
\subfigure[]{\includegraphics[width=0.75\columnwidth]{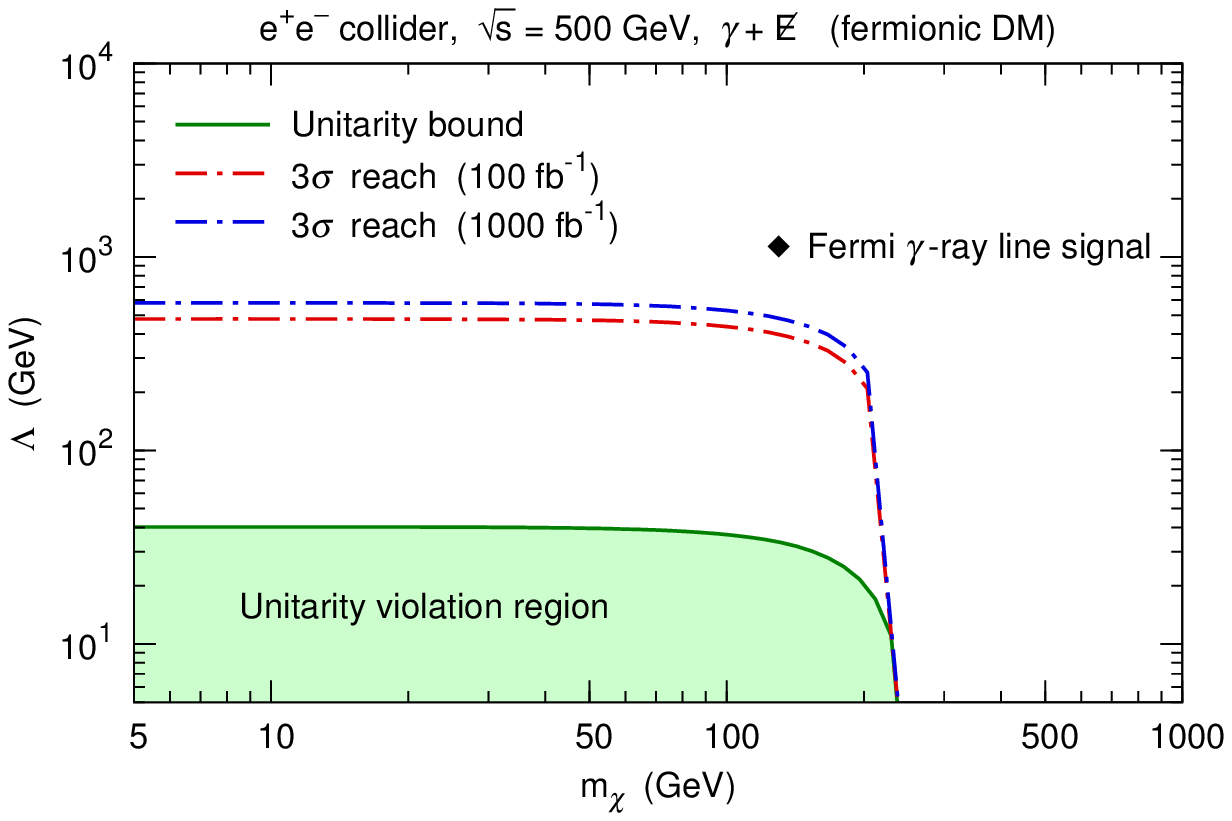}}
\subfigure[]{\includegraphics[width=0.75\columnwidth]{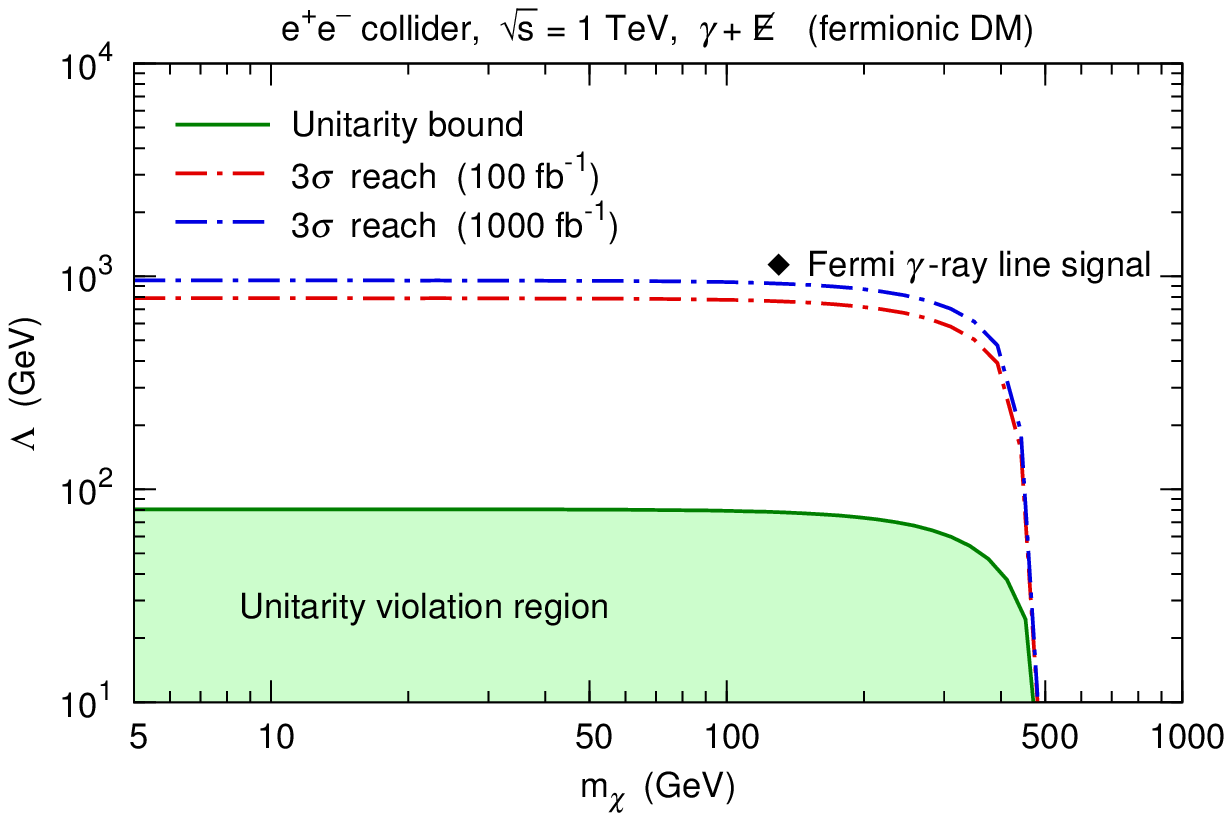}}
\subfigure[]{\includegraphics[width=0.75\columnwidth]{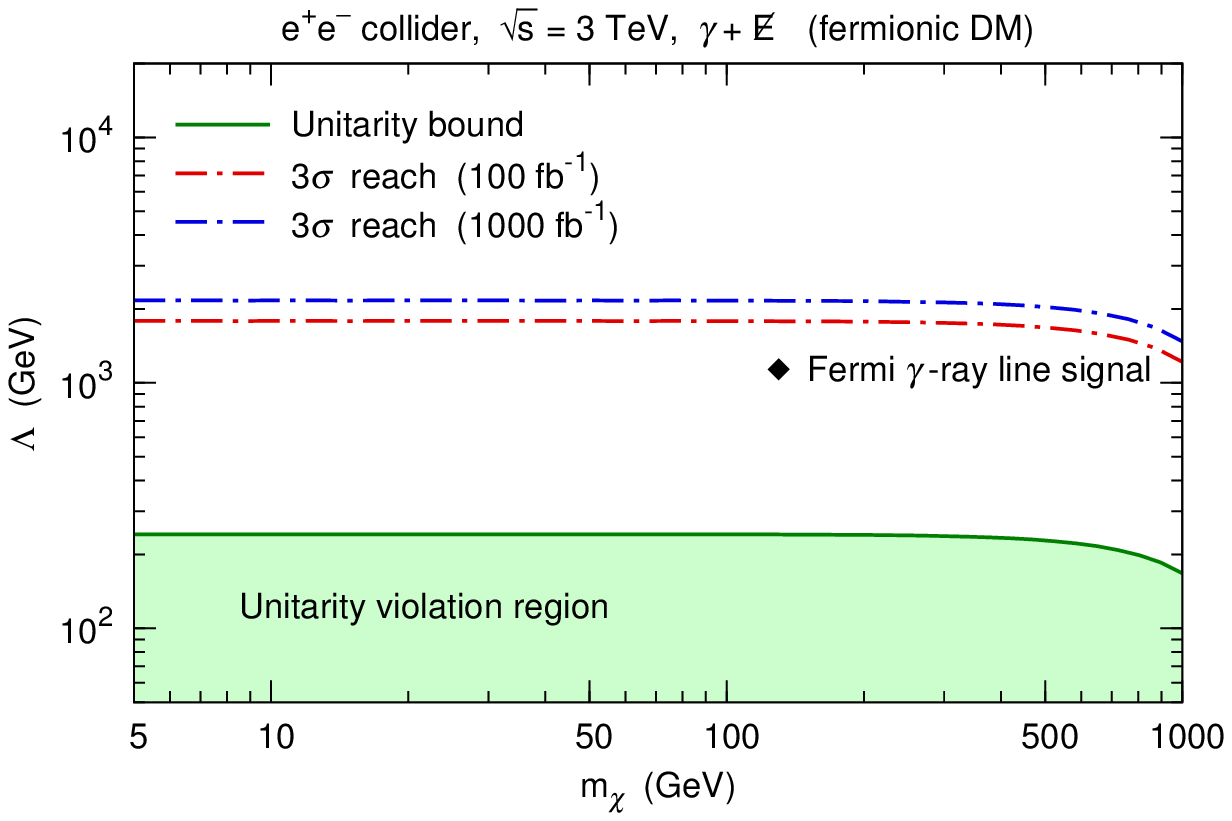}}
\caption{Unitarity bounds are shown on the fermionic DM production process $e^+e^-\to\chi\bar\chi\gamma$ at $e^+e^-$ colliders with $\sqrt{s}=250\,\GeV$, $500\,\GeV$, $1\,\TeV$, and $3\,\TeV$. The $3\sigma$ reaches are identical to those in Fig.~\ref{fig:reaches}.}
\label{fig:uni_FDM}
\end{figure*}

\begin{figure*}[!htbp]
\centering
\subfigure[]{\includegraphics[width=0.75\columnwidth]{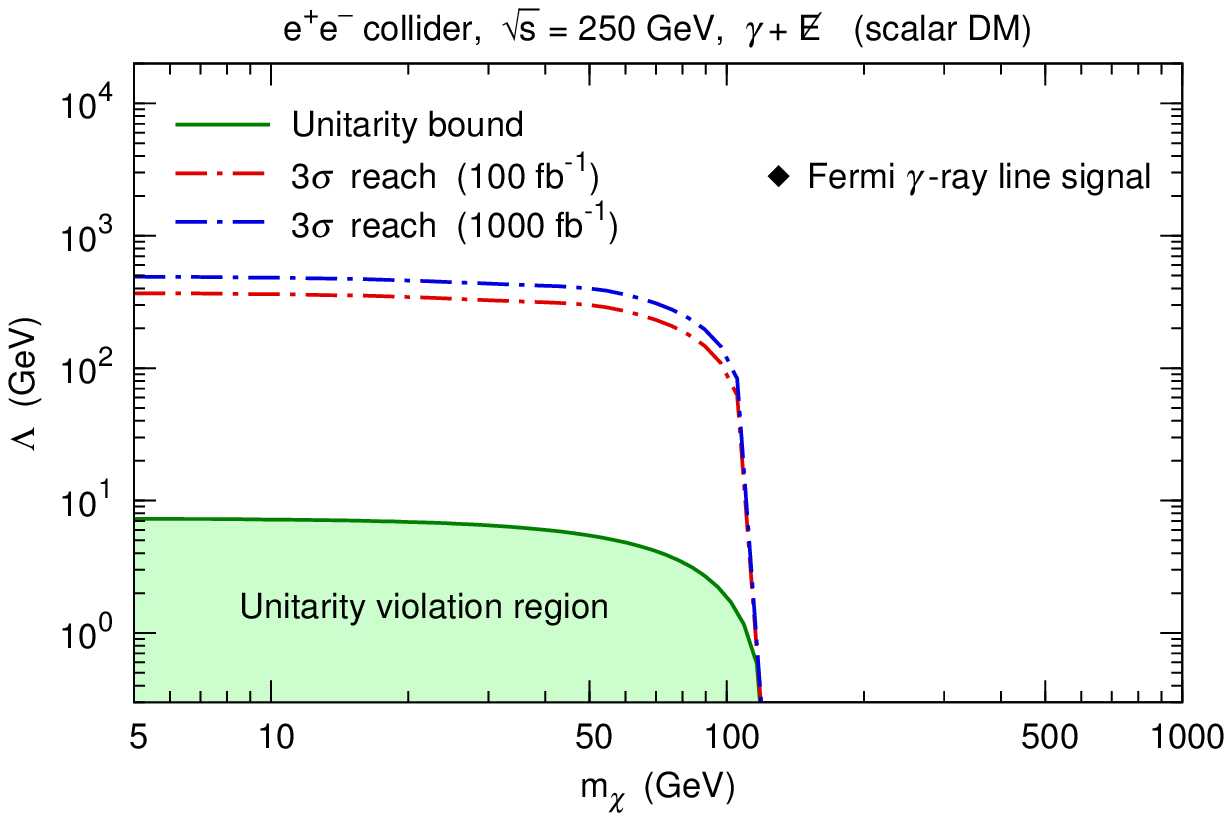}}
\subfigure[]{\includegraphics[width=0.75\columnwidth]{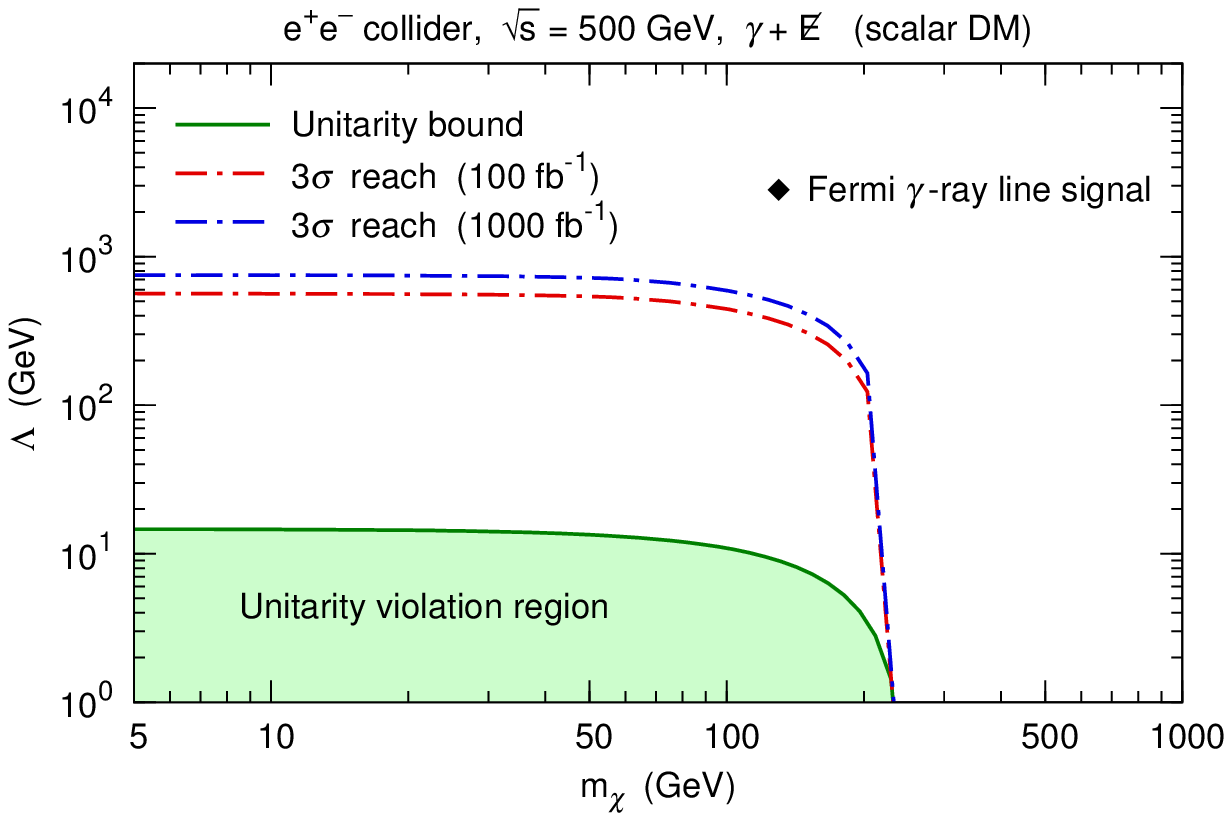}}
\subfigure[]{\includegraphics[width=0.75\columnwidth]{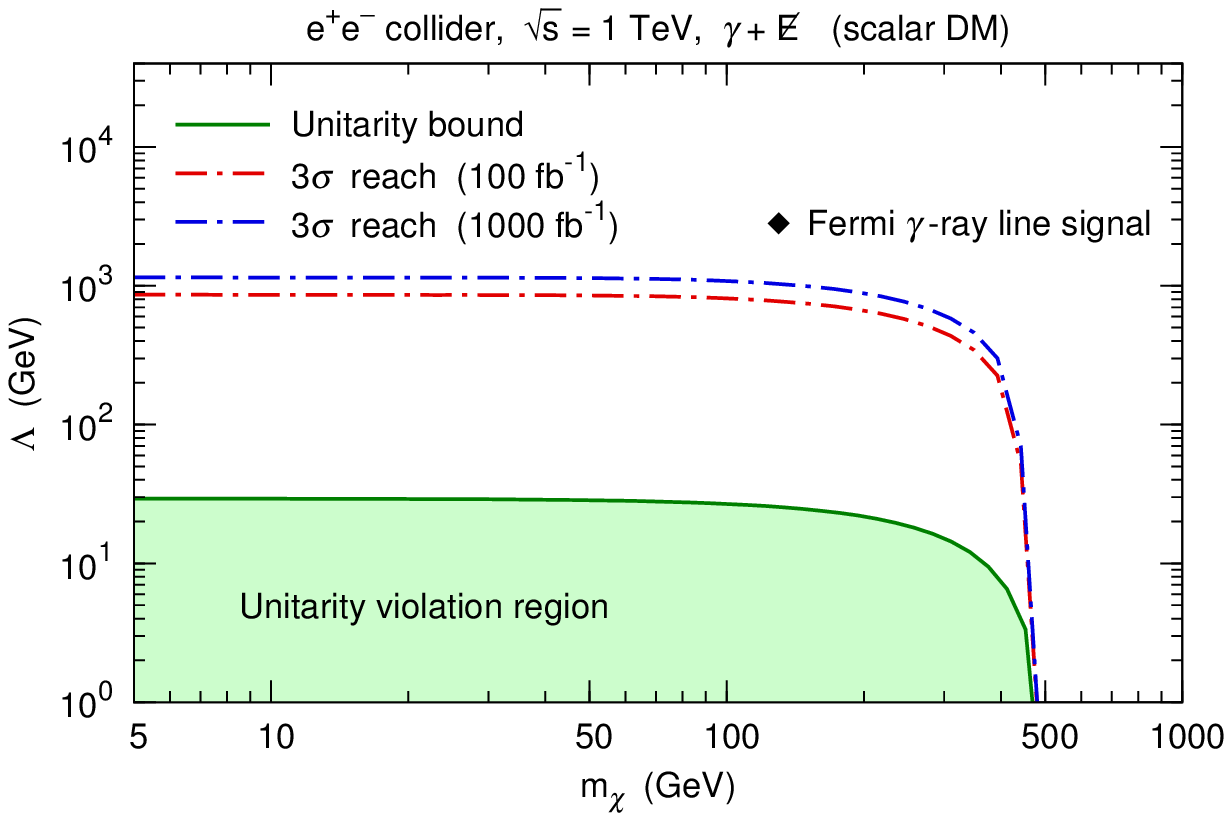}}
\subfigure[]{\includegraphics[width=0.75\columnwidth]{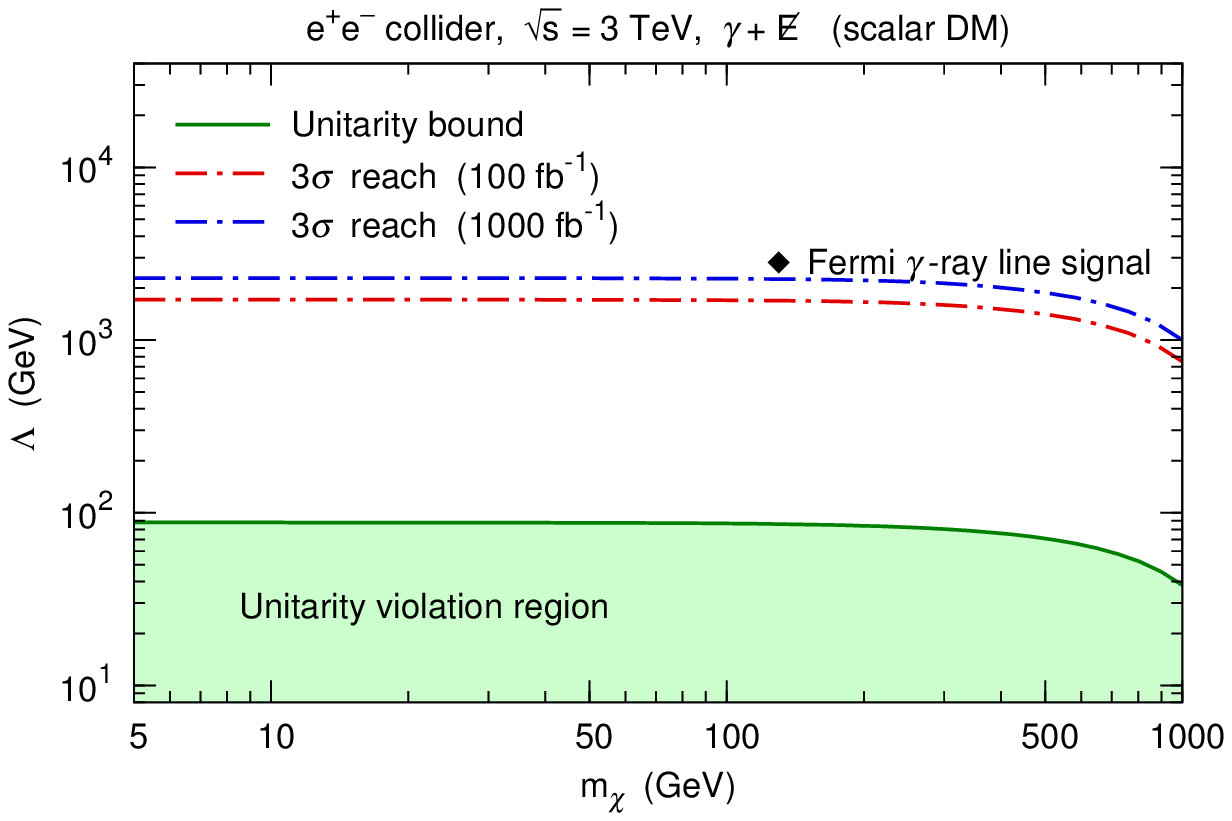}}
\caption{Same as Fig.~\ref{fig:uni_FDM} but for the scalar DM
production process $e^+e^-\to\chi\chi^*\gamma$.}
\label{fig:uni_SDM}
\end{figure*}

Then, we apply this general unitarity condition \eqref{ieq:b_inel} to the $2 \to 3$
DM production process $e^+e^-\to\chi\chi\gamma$.
The detailed calculation about $|b_j^\mathrm{inel}|$
can be found in Appendix~\ref{app:calc}.
At $e^+e^-$ colliders, $|b_0^\mathrm{inel}| \leq 1/2$
gives the most stringent unitarity bounds,
which have been shown in Figs.~\ref{fig:uni_FDM} and \ref{fig:uni_SDM}.
We find that all the $3\sigma$ reaches obtained above
lie far beyond the unitarity violation regions and pass this unitarity check.
The white regions between the $3\sigma$ reaches and the unitarity bounds
are the meaningful searching regions in the framework of the effective field theory.
This means that from the viewpoint of $S$-matrix unitarity
our effective operator treatment do not exceed its valid range.

It is remarkable that the unitarity condition Eq.~\eqref{ieq:b_inel} is derived
without any approximation. In Ref.~\cite{Dicus:2004rg},
a unitarity bound on the $2 \to n$ inelastic cross section
$\sigma_\mathrm{inel}(2 \to n)$ was obtained by using an approximate method as
\begin{equation}
\sigma_\mathrm{inel}(2 \to n) \leq \frac{4\pi}{s}.
\label{ieq:sigma_inel}
\end{equation}
We have compared the results given by Eq.~\eqref{ieq:sigma_inel}
with those given by our formula Eq.~\eqref{ieq:b_inel},
and find that their differences are negligibly small
for the processes considered here.

\section{Conclusions and discussions \label{sec:conc}}

In this work, we explore the sensitivity to the effective operators of DM and photons at TeV-scale $e^+e^-$ colliders.
The $\gamma + \missE$ signature is used to search for
the DM production process $e^+e^-\to\chi\chi\gamma$. Comparing with the indirect detection facilities, $e^+e^-$ colliders can offer a unique way to probe that for the signal of DM particles. With a $100\,\ifb$ dataset, the potential Fermi $\gamma$-ray line signal for the fermionic DM can be tested at a $3\,\TeV$ collider, though the scalar DM searching would be challenging.

Polarized beams at $e^+e^-$ colliders can suppress the SM background events efficiently and can enhance the production rate of the signals considerably. It is found that using the polarized beams is roughly equivalent to collecting 10 times of data. After considering a realistic polarization configuration,
we find that the potential Fermi signal can be tested by using a $2000\,\ifb$ dataset at $\sqrt{s}=1\,\TeV$ if the DM particle is a fermion, and a $1000\,\ifb$ dataset at a $3\,\TeV$ collider if the DM particle is a scalar.

To examine the validity of the effective operator approach,
we derive a general unitarity condition for $2 \to n$ processes
and apply it to the DM searching process $e^+e^-\to\chi\chi\gamma$. We find that our effective operator treatment can be justified from the viewpoint of $S$-matrix unitarity condition. The $3\sigma$ reaches we obtained are valid since they are far beyond the unitarity violation regions.

\begin{acknowledgments}
The authors thank Xiao-Jun Bi for helpful discussions. This work is supported by the Natural Science Foundation of China under Grants No. 11105157 and No. 11175251.
\end{acknowledgments}

\setcounter{equation}{0}
\renewcommand{\theequation}{\arabic{section}.\arabic{equation}}%

\appendix

\section{Detailed calculations for the unitarity bound \label{app:calc}}

In this appendix, we give detailed calculations of $|b_j^\mathrm{inel}|$
for the $2 \to 3$ DM production process $e^+e^-\to\chi\chi\gamma$.
For a $2 \to 3$ process, we can factorize Eq.~\eqref{eq:b_inel_def} to be
\begin{equation}
|b_j^\mathrm{inel}|^2 =
\frac{1}{64\pi} \int_{-1}^1 d\cos\theta_{\alpha\beta}
P_j(\cos\theta_{\alpha\beta}) G(\theta_{\alpha\beta}),
\end{equation}
where
\begin{eqnarray}
G(\theta_{\alpha\beta}) &\equiv &
\int d\Pi_{\gamma_3} \sum_{\text{spins of }\gamma_3}
\mathcal{M}_{\beta\to\gamma_3}^* \mathcal{M}_{\alpha\to\gamma_3}
\nonumber\\
&&\times(2\pi)^4 \delta^{(4)}(p_\alpha-p_{\gamma_3})
\nonumber\\
&=&\int \frac{d^3 k_3}{(2\pi)^3 2k_3^0} \frac{d^3 k_4}{(2\pi)^3 2k_4^0}
\frac{d^3 k_5}{(2\pi)^3 2k_5^0}
\nonumber\\
&&\times(2\pi)^4 \delta^{(4)}(p_1+p_2-k_3-k_4-k_5)
\nonumber\\
&&\times\sum_{\text{spins of }\gamma_3}
\mathcal{M}_{\beta\to\gamma_3}^* \mathcal{M}_{\alpha\to\gamma_3}.
\label{eq:G}
\end{eqnarray}
Here we sum over the spin states of the final state $\gamma_3$
to include all the available states of the $2 \to 3$ process.
Note that the $2 \to 3$ cross section can be related to
$G(\theta_{\alpha\beta})$ through
\begin{equation}
\sigma(2 \to 3) = \frac{1}{2s} G(\theta_{\alpha\beta}=0),
\end{equation}
for massless incoming particles.

For the scalar DM, we consider the transitions of state
\begin{eqnarray}
\alpha\to\gamma_3 &:~& e^-_{\lambda}(p_1) + e^+_{\lambda'}(p_2)
\to \gamma(k_3) + \chi(k_4) + \chi^*(k_5),
\nonumber\\
\beta\to\gamma_3 &:~& e^-_{\lambda}(q_1) + e^+_{\lambda'}(q_2)
\to \gamma(k_3) + \chi(k_4) + \chi^*(k_5),
\nonumber\\
\end{eqnarray}
where $\lambda,\lambda'=\pm$ are the helicity eigenvalues
(``$+$'' for right-handed, ``$-$'' for left-handed).
The correspond amplitudes are
\begin{eqnarray}
i\mathcal{M}_{\alpha\to\gamma_3}
&=& -i\frac{4e}{\Lambda^2 s} \bar v_{\lambda'}(p_2) \gamma^\mu u_\lambda(p_1)
\nonumber\\
&&\times[(k_3 \cdot q) \varepsilon_\mu^*(k_3)
- k_{3\mu} q^\nu \varepsilon_\nu^*(k_3)],
\nonumber\\
(i\mathcal{M}_{\beta\to\gamma_3})^*
&=& i\frac{4e}{\Lambda^2 s}\bar u_\lambda(q_1)\gamma^\rho v_{\lambda'}({q_2})
\nonumber\\
&&\times[(k_3 \cdot q) \varepsilon_\rho(k_3)
- k_{3\rho} q^\sigma \varepsilon_\sigma(k_3)],
\end{eqnarray}
where $q=p_1+p_2=q_1+q_2$ and $\varepsilon_\mu (k_3)$ is
the polarization vector of the photon. Hence we have
\begin{equation}
\sum_{\text{spins of }\gamma_3}
\mathcal{M}_{\beta\to\gamma_3}^* \mathcal{M}_{\alpha\to\gamma_3}
=\frac{64\pi\alpha}{\Lambda^4 s^2} F(e^-_\lambda,e^+_{\lambda'}),
\end{equation}
where
\begin{eqnarray}
F(e^-_\lambda,e^+_{\lambda'}) &\equiv& - \bar v_{\lambda'}(p_2) \gamma^\mu
u_\lambda(p_1) \bar u_\lambda(q_1) \gamma^\rho v_{\lambda'}(q_2)
\nonumber\\
&&\quad\times[(k_3 \cdot q)^2 g_{\mu\rho} + s k_{3\mu} k_{3\rho}].
\label{eq:F}
\end{eqnarray}

In the Weyl representation, the Dirac spinors $u_\lambda(p)$ and
$v_\lambda(p)$ can be expressed by helicity states $\xi_\lambda(p)$:
\begin{eqnarray}
u_\lambda(p) &=& \left(
\begin{array}{c}
  \omega_{-\lambda}(p) \xi_\lambda(p)  \\
  \omega_\lambda(p) \xi_\lambda(p)  \\
\end{array}
\right),
\nonumber\\
v_\lambda(p) &=& \left(
\begin{array}{c}
  -\lambda \omega_\lambda(p) \xi_{-\lambda}(p)\\
   \lambda \omega_{-\lambda}(p) \xi_{-\lambda}(p)\\
 \end{array}
\right),
\end{eqnarray}
where $\omega_\lambda(p) = \sqrt{E + \lambda |\mathbf{p}|}$.
For the initial state $\alpha$ with momenta
\begin{equation}
p_1 = \frac{\sqrt{s}}{2}(1,0,0, \beta_e),\quad
p_2 = \frac{\sqrt{s}}{2}(1,0,0,-\beta_e),
\end{equation}
where $\beta_e\equiv\sqrt{1-4m_e^2/s}$, the corresponding helicity states are
\begin{eqnarray}
\xi_+ (p_1) &=& \left(
\begin{array}{c}
   1  \\
   0  \\
\end{array} \right),\quad
\xi_- (p_1) = \left(
\begin{array}{c}
   0  \\
   1  \\
\end{array} \right),
\nonumber\\
\xi_+ (p_2) &=& \left(
\begin{array}{c}
   0  \\
  -1  \\
\end{array} \right),\quad
\xi_- (p_2) = \left(
\begin{array}{c}
   1  \\
   0  \\
\end{array} \right).
\end{eqnarray}
For the initial state $\beta$ with momenta
\begin{eqnarray}
q_1 &=& \frac{\sqrt{s}}{2}(1, \beta_e \sin\theta_{\alpha\beta},
0, \beta_e \cos\theta_{\alpha\beta}),
\nonumber\\
q_2 &=& \frac{\sqrt{s}}{2}(1,-\beta_e \sin\theta_{\alpha\beta},
0,-\beta_e \cos\theta_{\alpha\beta}),
\end{eqnarray}
the corresponding helicity states are
\begin{eqnarray}
\xi_+ (q_1) &=& \left(
\begin{array}{r}
  \cos\dfrac{\theta_{\alpha\beta}}{2}  \\
  \sin\dfrac{\theta_{\alpha\beta}}{2}  \\
\end{array} \right),\quad
\xi_- (q_1) = \left(
\begin{array}{r}
 -\sin\dfrac{\theta_{\alpha\beta}}{2}  \\
  \cos\dfrac{\theta_{\alpha\beta}}{2}  \\
 \end{array} \right),
\nonumber\\
\xi_+ (q_2) &=& \left(
\begin{array}{r}
  \sin\dfrac{\theta_{\alpha\beta}}{2}  \\
 -\cos\dfrac{\theta_{\alpha\beta}}{2}  \\
\end{array} \right),\quad
\xi_- (q_2) = \left(
\begin{array}{r}
  \cos\dfrac{\theta_{\alpha\beta}}{2}  \\
  \sin\dfrac{\theta_{\alpha\beta}}{2}  \\
\end{array} \right).
\nonumber\\
\end{eqnarray}

Using these expressions, we can compute $F(e^-_\lambda,e^+_{\lambda'})$
according to Eq.~\eqref{eq:F}. The results are
\begin{eqnarray}
F(e^-_+,e^+_-) &=& \frac{1}{2} s^2 |\mathbf{k}_3|^2
\big[ (1 + \cos\theta_{\alpha\beta})(1 + \cos^2\theta_3)
\nonumber\\
&&\qquad\qquad + (1 - \cos\theta_{\alpha\beta}) \sin^2\theta_3 \cos 2\phi_3
\nonumber\\
&&\qquad\qquad + \sin\theta_{\alpha\beta}\sin 2\theta_3 \cos\phi_3
\nonumber\\
&&\qquad\qquad + i(1 - \cos\theta_{\alpha\beta}) \sin^2\theta_3 \sin 2\phi_3
\nonumber\\
&&\qquad\qquad + i\sin\theta_{\alpha\beta} \sin 2\theta_3 \sin\phi_3 \big],
\nonumber\\
F(e^-_-,e^+_+) &=& [F(e^-_+,e^+_-)]^*,
\nonumber\\
F(e^-_-,e^+_-) &=& 2s|\mathbf{k}_3|^2 m_e^2
\big( 2\cos\theta_{\alpha\beta} \sin^2\theta_3
\nonumber\\
&&\qquad\qquad\quad - \sin\theta_{\alpha\beta} \sin 2\theta_3\cos\phi_3 \big),
\nonumber\\
F(e^-_+,e^+_+) &=& F(e^-_-,e^+_-),
\label{eq:F_explicit}
\end{eqnarray}
where $\theta_3$ and $\phi_3$ are the zenith and the azimuthal angles
of the photon, respectively.
After the integration over $\phi_3$, the imaginary parts of
$F(e^-_+,e^+_-)$ and $F(e^-_-,e^+_+)$ vanish,
and they give the same $|b_j^\mathrm{inel}|^2$.
Due to helicity suppression, $F(e^-_-,e^+_-)$ and $F(e^-_+,e^+_+)$
are proportional to $m_e^2$, and their resulting
$|b_j^\mathrm{inel}|^2 \propto m_e^2/s$, which vanish for $s \gg m_e^2$.
Therefore, the essential unitarity bounds are given by the processes
where the initial $e^-$ and $e^+$ have different helicities.

For the fermionic DM, we consider the transitions of state
\begin{eqnarray}
\alpha\to\gamma_3 &:~& e^-_{\lambda}(p_1) + e^+_{\lambda'}(p_2)
\to \gamma(k_3) + \chi(k_4) + \bar\chi(k_5),
\nonumber\\
\beta\to\gamma_3 &:~& e^-_{\lambda}(q_1) + e^+_{\lambda'}(q_2)
\to \gamma(k_3) + \chi(k_4) + \bar\chi(k_5),
\nonumber\\
\end{eqnarray}
where we do not denote the helicity eigenvalues of DM particles,
which will be summed over in the following calculations.
The corresponding amplitudes are
\begin{eqnarray}
i\mathcal{M}_{\alpha\to\gamma_3}
&=& - \frac{4e}{\Lambda^3 s} \bar v_{\lambda'}(p_2) \gamma_\mu u_\lambda(p_1)
\bar u(k_4) \gamma_5 v(k_5)
\nonumber\\
&&\times \varepsilon^{\nu\mu\rho\sigma} k_{3\rho} q_\sigma
\varepsilon_\nu^*(k_3),
\nonumber\\
(i\mathcal{M}_{\beta\to\gamma_3})^*
&=& \frac{4e}{\Lambda^3 s}\bar u_\lambda(q_1)\gamma_\delta v_{\lambda'}({q_2})
\bar v(k_5) \gamma_5 u(k_4)
\nonumber\\
&&\times \varepsilon^{\gamma\delta\alpha\beta} k_{3\alpha} q_\beta
\varepsilon_\gamma (k_3).
\end{eqnarray}
Then we have
\begin{equation}
\sum_{\text{spins of }\gamma_3}
\mathcal{M}_{\beta\to\gamma_3}^* \mathcal{M}_{\alpha\to\gamma_3}
=\frac{256\pi\alpha}{\Lambda^6 s^2} (k_4 \cdot k_5 + m_\chi^2)
F(e^-_\lambda,e^+_{\lambda'}),
\end{equation}
where $F(e^-_\lambda,e^+_{\lambda'})$ is the same quantity
defined in Eq.~\eqref{eq:F}, and the results in \eqref{eq:F_explicit}
can also be used for the case of the fermionic DM.

Now let us make explicit the phase space integration in Eq.~\eqref{eq:G}.

Using $\int dk_{35}^0 \delta(s_{35} - k_{35}^2)
= \big(2\sqrt{s_{35} + |\mathbf{k}_{35}|^2}\big)^{-1}$
with $k_{35}=k_3+k_5$, we can split the 3-body phase space integration
into two 2-body phase space integrations:
\begin{eqnarray}
\int {d{\Phi ^{(3)}}}
&\equiv& \int \frac{d^3 k_3}{(2\pi)^3 2k_3^0} \frac{d^3 k_4}{(2\pi)^3 2k_4^0}
\frac{d^3 k_5}{(2\pi)^3 2k_5^0}
\nonumber\\
&&\times (2\pi)^4 \delta^{(4)}(p_1+p_2-k_3-k_4-k_5)
\nonumber\\
&=& \int d{s_{35}} d^4 k_{35} \delta (s_{35}-k_{35}^2)
\delta^{(4)}(k_{35}-k_3-k_5)
\nonumber\\
&&\times \frac{d^3 k_3}{(2\pi)^3 2k_3^0} \frac{d^3 k_4}{(2\pi)^3 2k_4^0}
\frac{d^3 k_5}{(2\pi)^3 2k_5^0}
\nonumber\\
&&\times (2\pi)^4 \delta^{(4)}(p_1+p_2-k_3-k_4-k_5)
\nonumber\\
&=& \int \frac{d s_{35}}{2\pi} d\Phi_1^{(2)} d\Phi_2^{(2)},
\end{eqnarray}
where
\begin{equation}
d\Phi_1^{(2)} \equiv \frac{d^3 k_4}{(2\pi)^3 2k_4^0}
\frac{d^3 k_{35}}{(2\pi)^3 2k_{35}^0}
(2\pi)^4 \delta^{(4)}(p_1+p_2-k_4-k_{35}),
\end{equation}
\begin{equation}
d\Phi_2^{(2)} \equiv \frac{d^3 k_3}{(2\pi)^3 2k_3^0}
\frac{d^3 k_5}{(2\pi)^3 2k_5^0}
(2\pi)^4 \delta^{(4)}(k_{35}-k_3-k_5).
\end{equation}

According to 2-body kinematics,
in the center-of-mass frame of $p_1$ and $p_2$,
\begin{equation}
k_{35}^0 = \frac{s+s_{35}-m_\chi^2}{2\sqrt{s}},\quad
k_4^0 = \frac{s+m_\chi^2-s_{35}}{2\sqrt{s}},
\end{equation}
and
\begin{equation}
|{{\mathbf{k}}_4}| = \frac{1}{{2\sqrt s }}\sqrt {\left[ {s
- {{(\sqrt {{s_{35}}}  + {m_\chi })}^2}} \right]\left[ {s
- {{(\sqrt {{s_{35}}}  - {m_\chi })}^2}} \right]},
\end{equation}
while in the center-of-mass frame of $k_3$ and $k_5$,
\begin{equation}
\tilde k_{35}^\mu  = (\sqrt {s_{35}} ,0,0,0)
~~\text{and}~~
\tilde k_3^0 = \frac{{{s_{35}} - m_\chi ^2}}{{2\sqrt {s_{35}} }}.
\end{equation}
Expressing $\mathbf{k}_3$ and $\mathbf{k}_4$ as
\begin{equation}
\mathbf{k}_3 = |\mathbf{k}_3|(\sin\theta_3\cos\phi_3,
\sin\theta_3\sin\phi_3,\cos\theta_3)
\end{equation}
and
\begin{equation}
\mathbf{k}_4 = |\mathbf{k}_4|(\sin\theta_4,0,\cos\theta_4)
=-\mathbf{k}_{35},
\end{equation}
and using the Lorentz invariant property
$k_{35} \cdot k_3 = \tilde k_{35} \cdot \tilde k_3$, we can obtain
\begin{equation}
|\mathbf{k}_3| = \frac{s_{35}-m_\chi^2}
{2[k_{35}^0 + |\mathbf{k}_4|A(\theta_4,\theta_3,\phi_3)]},
\end{equation}
where $A(\theta_4,\theta_3,\phi_3) \equiv
\sin\theta_4 \sin\theta_3 \cos\phi_3 + \cos\theta_4\cos\theta_3$.
Due to $\mathbf{k}_3+\mathbf{k}_4+\mathbf{k}_5=0$, we have
\begin{equation}
|\mathbf{k}_5|^2 = |\mathbf{k}_3|^2 + |\mathbf{k}_4|^2
+ 2 |\mathbf{k}_3| |\mathbf{k}_4| A(\theta_4,\theta_3,\phi_3),
\end{equation}
and
\begin{equation}
\frac{\partial\sqrt{|\mathbf{k}_5|^2 + m_\chi^2}}{\partial |\mathbf{k}_3|}
= \frac{|\mathbf{k}_3| + |\mathbf{k}_4| A(\theta_4,\theta_3,\phi_3)}
{\sqrt{|\mathbf{k}_5|^2 + m_\chi^2}}.
\end{equation}
Then we can simplify the integrations over $\Phi_1^{(2)}$ and $\Phi_2^{(2)}$:
\begin{eqnarray}
\int d\Phi_1^{(2)}
&=& \int \frac{|\mathbf{k}_4|^2 d|\mathbf{k}_4| d\cos\theta_4}
{8\pi k_4^0 k_{35}^0}
\nonumber\\
&&\times \delta \left(p_1^0 + p_2^0 - \sqrt{|\mathbf{k}_4|^2 + m_\chi^2}
 - \sqrt{|\mathbf{k}_4|^2 + s_{35}} \right)
\nonumber\\
&=& \int \frac{|\mathbf{k}_4|^2 d\cos\theta_4}{8\pi k_4^0 k_{35}^0}
\left(\frac{|\mathbf{k}_4|}{k_4^0}
+ \frac{|\mathbf{k}_4|}{k_{35}^0} \right)^{-1}
\nonumber\\
&=& \frac{|\mathbf{k}_4|}{8\pi\sqrt{s}} \int d\cos\theta_4,
\end{eqnarray}
\begin{eqnarray}
\int d\Phi_2^{(2)}
&=& \int \frac{|\mathbf{k}_3|^2 d|\mathbf{k}_3| d\cos\theta_3 d\phi_3}
{16 \pi^2 k_3^0 k_5^0}
\nonumber\\
&&\times \delta \left( k_{35}^0 - |\mathbf{k}_3|
 - \sqrt{|\mathbf{k}_5|^2 + m_\chi^2} \right)
\nonumber\\
 &=& \int \frac{|\mathbf{k}_3|^2 d\cos\theta_3 d\phi_3}
{16 \pi^2 k_3^0 k_5^0}
\nonumber\\
&&\times \left[ \frac{|\mathbf{k}_3|}{k_3^0}
+ \frac{|\mathbf{k}_3|}{k_5^0} + \frac{|\mathbf{k}_4|}{k_5^0}
A(\theta_4,\theta_3,\phi_3) \right]^{-1}
\nonumber\\
&=& \frac{1}{8\pi^2 (s_{35} - m_\chi^2)}
 \int d\cos\theta_3 d\phi_3 |\mathbf{k}_3|^2.
\end{eqnarray}
Therefore, $G(\theta_{\alpha\beta})$ can be expressed as
\begin{eqnarray}
G(\theta_{\alpha\beta})
&=& \int \frac{d s_{35}}{2\pi} d\Phi_1^{(2)} d\Phi_2^{(2)}
\sum_{\text{spins of }\gamma_3}
\mathcal{M}_{\beta\to\gamma_3}^* \mathcal{M}_{\alpha\to\gamma_3}
\nonumber\\
&=& \frac{1}{128\pi^4}\int_{m_\chi^2}^{(\sqrt{s}-m_\chi)^2}
 d s_{35} \int_0^\pi d\theta_4 \sin\theta_4
\nonumber\\
&&\times \int_0^\pi d\theta_3 \sin\theta_3 \int_0^{2\pi} d\phi_3
\frac{|\mathbf{k}_4| |\mathbf{k}_3|^2}{\sqrt{s}(s_{35}-m_\chi^2)}
\nonumber\\
&&\times \sum_{\text{spins of }\gamma_3}
\mathcal{M}_{\beta\to\gamma_3}^* \mathcal{M}_{\alpha\to\gamma_3}.
\end{eqnarray}
When computing $G(\theta_{\alpha\beta})$ for the fermionic DM,
we also need the expression
\begin{equation}
k_4 \cdot k_5 = k_4^0 k_5^0 + |\mathbf{k}_4|^2
+ |\mathbf{k}_4| |\mathbf{k}_3|A(\theta_4,\theta_3,\phi_3),
\end{equation}
where $k_5^0=\sqrt{|\mathbf{k}_5|^2+m_\chi^2}$.

Now we have all the expressions which are needed to calculate
$|b_j^\mathrm{inel}|$ for the process $e^+e^-\to\chi\chi\gamma$.
We perform the integrations by numerical methods
and obtain the unitarity bounds corresponding to $|b_0^\mathrm{inel}|\leq 1/2$.
They have been shown in Figs.~\ref{fig:uni_FDM} and \ref{fig:uni_SDM}.

\end{document}